\documentclass[aps,showpacs,twocolumn,superscriptaddress]{revtex4}
\usepackage{amsmath}
\usepackage{amssymb}
\usepackage{amscd}
\usepackage{wasysym}
\usepackage[ansinew]{inputenc}
\usepackage[T1]{fontenc}
\usepackage{ae,aecompl}
\usepackage{subfigure}

    \usepackage[dvips]{graphicx}
    \DeclareGraphicsExtensions{.eps}

\newcommand{\ri}{ i }

\newcommand{\be}{\begin{eqnarray}}
\newcommand{\ee}{\end{eqnarray}}
\newcommand{\nn}{\nonumber}

\newcommand{\geapprox}{\apprge}
\newcommand{\ket}[1]{|#1\rangle}
\newcommand{\bra}[1]{\left\langle#1\right|}

\renewcommand{\vec}[1]{\mathbf{#1}}

\newcommand{\HH}{\mathcal{H}}
\newcommand{\Szwo}{\mathcal{S}^2}

\renewcommand{\Im}{{\rm Im}}
\newcommand{\Dop}{\mathcal{D}}
\newcommand{\abl}[1]{\frac{\partial}{\partial #1}}
\newcommand{\tr}{{\rm tr}}

\begin{document}
\bibliographystyle{apsrev}

\title{Beyond mean-field dynamics of small Bose-Hubbard systems based on the number-conserving phase space approach}

\author{F. Trimborn}
\affiliation{Fachbereich Physik, Technische Universit{\"a}t Kaiserslautern,
D-67653 Kaiserslautern, Germany}
\affiliation{Institut f\"ur mathematische Physik, TU Braunschweig, D--38106 Braunschweig, Germany}
\author{D. Witthaut}
\email{dirk.witthaut@nbi.dk}
\affiliation{Fachbereich Physik, Technische Universit{\"a}t Kaiserslautern,
D-67653 Kaiserslautern, Germany}
\affiliation{QUANTOP, Niels Bohr Institute, University of Copenhagen, 
DK--2100 Copenhagen, Denmark}
\author{H. J. Korsch}
\affiliation{Fachbereich Physik, Technische Universit{\"a}t Kaiserslautern,
D-67653 Kaiserslautern, Germany}

\date{\today }

\begin{abstract}
The number-conserving quantum phase space description of the Bose-Hubbard model is 
discussed for the illustrative case of two and three modes, as well as the 
generalization of the two-mode case to an open quantum system.
The phase-space description based on generalized $SU(M)$ coherent states yields a 
Liouvillian flow in the macroscopic limit, which can be efficiently simulated using 
Monte Carlo methods even for large systems. We show that this description 
clearly goes beyond the common mean-field limit. In particular it resolves 
well-known problems where the common mean-field approach fails, like the description 
of dynamical instabilities and chaotic dynamics. Moreover, it provides a valuable tool
for a semi-classical approximation of many interesting quantities, which depend on higher 
moments of the quantum state and are therefore not accessible within the common approach.
As a prominent example, we analyse the depletion and heating of the condensate.
A comparison to methods ignoring the fixed particle number shows that in this case 
artificial number fluctuations lead to ambiguities and large deviations even for quite 
simple examples.
\end{abstract}

\maketitle


\section{Introduction}

The physics of ultracold  atoms in optical lattices has made an enormous 
progress in the last decade, since it is an excellent model system for a 
variety of fields such as nonlinear dynamics or condensed matter physics.
The dynamics of bosonic atoms can be described by the celebrated Bose-Hubbard 
Hamiltonian, which is a paradigmatic model for the study of strongly 
correlated many-body quantum systems \cite{Fish89b}. 
Such systems are hard to deal with theoretically since the dimension of the
respective Hilbert space increases exponentially both in the particle number
and in the number of lattice sites. Therefore, approximations to the dynamics
are of considerable interest.

However, things can become surprisingly simple if the atoms undergo Bose-Einstein
condensation. In many situations, the mesoscopic dynamics of a Bose-Einstein 
condensate (BEC) is extremely well described by the (discrete) Gross-Pitaevskii 
equation (GPE) for the macroscopic wavefunction of the condensate (see, e.g., 
\cite{Peth08}).
This mean-field description is often referred to as 'classical' since the macroscopic 
limit $N \rightarrow \infty$ of the dynamics in second quantization is formally equivalent 
to the limit $\hbar \rightarrow 0$ which yields classical mechanics as the limit of 
single particle quantum mechanics. 
In practice, this approximation is frequently derived within a Bogoliubov approach, 
factorizing the expectation values of products of operators into the product 
of expectation values. As a consequence there is no obvious indicator to quantify 
the errors or to specify the scope of validity. Moreover, many interesting observables
which involve higher moments of quantum state are not accessible within this approximation.
A more detailed overview over established methods and a comparison to the approach 
presented here will be given in Sec.~\ref{sec-comparison}.

The phase space formulation of quantum mechanics is nearly as old as the 
theory itself and has a wide range of applications, especially in quantum optics. 
Beneath the illustrative insight into the dynamics provided by this formulation, many 
techniques and methods were developed in this context. 
However, only a small fraction of the literature is dedicated to systems with 
intrinsic symmetries like the Bose-Hubbard Hamiltonian. In this case the 
dynamical group is isomorphic to the special unitary group $SU(M)$, with $M$ being the 
number of lattice sites, reflecting the conservation of the total particle number.    
A general algorithm to construct phase space distribution functions for systems
whose dynamical group has the structure of an arbitrary Lie group such as $SU(M)$ has been 
developed only nine years ago \cite{Brif98b}. 
It is based on the concept of generalized coherent states introduced by 
Gilmore \cite{Zhan90} and Perelomov \cite{Pere86}. Starting from these states we have 
surveyed the mathematical foundations of the number conserving phase space description 
and derived the exact evolution equations for the Husimi $Q$-function and the 
Glauber-Sudarshan $P$-function of the $M$-site Bose-Hubbard model in a preceding paper 
\cite{07phase}. 
One important consequence of the use of $SU(M)$ coherent states is the different topology 
of the phase space, which is now isomorphic to the $2M-2$ dimensional Bloch sphere and
therefore compact. Moreover, the phase space description based on $SU(M)$ coherent states
allows for a deeper analysis of the many-particle-mean-field correspondence, since 
these states are of high physical significance. As they are equivalent to the fully condensed
product states, they provide an excellent tool to describe and study derivations from the 
macroscopic state which also determine the region validity of the mean-field approximation.

In this paper we will illustrate the methods originally developed for the $M$-site system \cite{07phase}
for the instructive case of small Bose-Hubbard systems. 
Therefore we consider a two-mode Bose Hubbard model
\be
  \hat H &=& 
  -\Delta(t)  \left( \hat a_1^\dagger \hat a_2 +  \hat a_2^\dagger \hat a_1   \right)
   + \epsilon _2 \hat a_2^\dagger \hat a_2  + \epsilon_1 \hat a_1^\dagger \hat a_1  \nn \\
   && \qquad + \frac{U}{2} \left( \hat a_1^{\dagger 2} \hat a_1^2 
            + \hat a_2^{\dagger 2} \hat a_2^2  \right),
    \label{eqn-hami-bh}
\ee
and its generalization which takes into account elastic collisions with background 
gas leading to phase noise described by a master equation. Such a system can 
be realized experimentally by confining a BEC in a double-well trap \cite{Albi05,Gati06,Schu05b,Foll07}. Moreover, we also consider the case of 
an explicitly time-dependent tunneling-element $\Delta$, which leads to mixed 
regular-chaotic mean-field dynamics.  This can be easily implemented in an 
experimental setup via a variation of the intensity of the laser forming the 
optical potential.  
However, this model can also describe the fundamental phenomena of two weakly 
coupled BECs in more general setups \cite{Orze01} and Landau-Zener transitions
between different Bloch bands \cite{Wu00,06zener_bec}. Early theoretical studies of 
the system dynamics without an external influence and a time-dependant driving were 
reported in \cite{Milb97,Smer97,Ragh99}.
Furthermore, we use the methods presented here to analyse the interesting and yet not completely 
understood case $M=3$, i.e. the three-mode Bose-Hubbard system:
\be
  \hat H &=& -\Delta_{12}  \left( \hat a_1^\dagger \hat a_2 +  \hat a_2^\dagger \hat a_1  \right) 
    -\Delta_{23}  \left( \hat a_2^\dagger \hat a_3 +  \hat a_3^\dagger \hat a_2  \right) \nn\\
    && + \sum_{j=1}^3 \epsilon_j \hat a_j^\dagger \hat a_j 
    + \frac{U}{2} \sum_{j=1}^3 \hat a_j^{\dagger 2} \hat a_j^2 .
      \label{eqn-hami-bh3}
\ee
The realization of such a linear triple-well trap for a BEC is not as straightforward 
as in the case of a double-well trap, but nevertheless possible in optical setups
\cite{Este08} or on an atom-chip. Besides it is a prominent model system because 
the mean-field dynamics is classically chaotic (see, e.g. 
\cite{Buon03,Fran03,05level3,Moss06}).

In particular, this paper is organised as follows: In the sections
\ref{sec-algebra} and \ref{sec-phasespace}, we briefly recall the main results 
from the number conserving phase space approach and provide some illustrative examples.
Then follows an analysis of the exact dynamics of the 
quasi probability distributions in Sec.~\ref{sec-dynamics}.
Here, we are especially interested in the macroscopic 
limit, since the evolution equations converge to a classical Liouville equation for 
large particle numbers. This result enables us to go beyond the usual mean-field 
limit which considers only point-distributions and to enhance the region of applicability 
of the approximation to arbitrary initial states. 
To emphasize this, we study in Sec.~\ref{sec-breakdown} the behaviour around dynamical 
instabilities where the common mean-field approach is known to fail badly. 
In the case of an isolated instability 
this effect is known as the breakdown of mean-field. Here, we show that the Liouville dynamics 
not only resolves this breakdown due to an isolated instability, but also provides
an ideal tool to describe systems where the mean-field dynamics faces large chaotic regions, 
as for example in a driven two mode system.

Having demonstrated the considerable advantages of the Liouville description, we are now able to 
address novel questions, like the dynamics of the three-mode system in Sec.~\ref{sec-beyond}.
While being conceptually quite simple and much easier to implement than other methods 
(cf. e.g. \cite{Cast97,Cast98}) the Liouville
description provides a valuable tool to estimate the depletion of the condensate mode 
and to infer the nature of the many-body quantum state. The comparison to exact numerical 
results show clear agreement with the results of the Liouville equation while the common mean-field approach fails.

Since the methods developed in \cite{07phase} are not restricted to the Bose-Hubbard model, but 
can be applied to every problem within the same symmetry group, there is a large variety of 
possible applications. As one example, we discuss the heating of a two-mode BEC resulting from 
elastic scattering with the background gas in Sec.~\ref{sec-heating}. 
While our methods predict the proper results, namely 
a quasi-classical set of Langevin equations in the mean-field approximation, they allow to go
even further and predict many interesting effects such as the decay of the coherence factor
and the increase of the variances, which are not accessible within the usual mean-field limit. This again underlines the benefit from the Liouville description and the distinction to common mean-field approaches.

In Sec.~\ref{sec-comparison} we then give a short overview over different approaches to derive 
and extend a mean-field description for the systems studied here.
Especially, we compare our approach based on the dynamical symmetry
and therefore explicitly including the conservation of the particle number to common phase space methods (cf. e.g. \cite{Stee98,Jain04}). One important result is that calculations based on a $U(1)$ symmetry breaking description using common Glauber coherent states show deviations from the exact dynamics which are significantly larger and appear on a shorter time scale compared to the number conserving approach. 

\section{Algebraic structure and generalized coherent states}    
\label{sec-algebra}

The number conserving phase space description of the Bose-Hubbard dynamics
has been introduced in \cite{07phase}, starting 
from the generalized coherent states of Gilmore \cite{Zhan90} and Perelomov 
\cite{Pere86}. Here we will only recall the main results for the special case 
of two and three modes for the sake of completeness.

Generalized coherent states for arbitrary dynamical Lie groups are defined by the 
action of a translation operator onto a reference state. For instance, the celebrated 
Glauber coherent states for the Heisenberg-Weyl group $H_4$ are defined by the relation
$\ket{\alpha} = \hat D(\alpha) \ket{0} = e^{\alpha \hat a^\dagger - \alpha^* \hat a} \ket{0}$,
where the reference state is just the vacuum state. The parameter space of the translation 
operators $\hat D(\alpha)$ and thus the space of the coherent states is isomorphic to the 
classical phase space $\mathbb{C}\simeq \mathbb{R}\times \mathbb{R}$. 

\subsection{The two-mode case}

The situation is different for a BEC in a double well trap since the dynamical group 
is now $SU(2)$ spanned by the angular momentum operators 
\be
  \hat J_x &=& \frac{1}{2} 
      \left( \hat a_1^\dagger \hat a_2  + \hat a_2^\dagger \hat a_1 \right), \nn \\
  \hat J_y &=& \frac{i}{2} 
      \left( \hat a_1^\dagger \hat a_2  - \hat a_2^\dagger \hat a_1 \right), 
      \label{eqn-angular-op} \\
  \hat J_z &=& \frac{1}{2} 
      \left( \hat a_2^\dagger \hat a_2  - \hat a_1^\dagger \hat a_1 \right). \nn   
\ee 
The Hamiltonian (\ref{eqn-hami-bh}) then can be rewritten  as
\be
  \hat H =  -2 \Delta \hat J_x + 2\epsilon \hat J_z   + U \hat J_z^2
  \label{eqn-hamiltonian-2level}
\ee
with $\epsilon = \epsilon_2 - \epsilon_1$ up to a constant term \cite{Milb97,Smer97,Vard01b,Fran00}.
The Casimir operator $\hat J^2 = \hat N/2 \,(\hat N/2 + 1)$ reveals the relation between the
algebraic structure and the total particle number $\hat N=\hat a_1^\dagger \hat a_1 + \hat a_2^\dagger \hat a_2$. 

Generalization of the concept of the translation operator to the $su(2)$ algebra yields an 
rotation operator, such that the $SU(2)$ or so called Bloch coherent states are defined as follows:
\be 
  \ket{\theta,\phi} &=& \hat R(\theta,\phi) \ket{N,0}  \label{eqn-blochstate}  \\
   = && \!\!\!\!\!\!\!\!  e^{-i \theta (\hat J_x \sin \phi - \hat J_y \cos \phi)} \ket{N,0} \nn \\
   = && \!\!\!\!\!\!\!\!  \sum_{n_1+n_2=N} \binom{N}{n_2}^{\frac 1 2}
      \cos\big({\scriptstyle \frac{\theta}{2}} \big)^{n_1} 
      \sin \big({\scriptstyle \frac{\theta}{2}} \big)^{n_2} 
        e^{-i n_2 \phi} \ket{n_1,n_2}. \nn
\ee  
The group of rotations $\hat R(\theta,\phi)$ and therefore the parameter space of coherent 
states can be described by two angles $0\leq \theta \leq \pi, 0\leq \phi < 2 \pi$ and is thus 
isomorphic to a Bloch sphere $\Szwo$.  
In physical terms, this can be understood as follows: The $z$-component of the Bloch 
vector describes the population imbalance of the two modes, while the polar angle 
represents the relative phase. 
Since this variable is cyclic and not defined if all particles are in a single well 
(i.e. at the poles), the topology is clearly that of a sphere.

Note that the $SU(M)$ coherent states are equivalent to the product states, 
i.e. they represent a pure condensate \cite{07phase,Buon08} For two modes, $M=2$, 
this relation simply reads
\be \label{eqn_productsu2}
  \ket{\theta,\phi} = \frac{1}{\sqrt{N!}} \left( \cos({ \scriptstyle \frac{\theta}{2}}) 
    a_1^\dagger + \sin({ \scriptstyle \frac{\theta}{2}}) 
    e^{-i \phi} a_2^\dagger \right)^N \ket{0,0}.
\ee
This representation of $SU(2)$ coherent states reveals a parametrization in more physical 
terms $p,q$, where  $q=\phi$ is equal to the relative phase between the modes and 
$p = \cos^2(\theta/2)$ denotes the population of the second mode.

\subsection{The three-mode case}

For the three-mode system the dynamical group is isomorphic to $SU(3)$ which is spanned by
the eight generators
\be
& \hat X_1=\hat a_1^\dagger \hat a_1 - \hat a_2^\dagger \hat a_2, \nn \\
& \hat X_2 = \frac 1 3 \left(\hat a_1^\dagger \hat a_1 + \hat a_2^\dagger \hat a_2 - 2 \hat a_3^\dagger \hat a_3\right), \nn \\
& \hat Y_k = i \left(\hat a_k^\dagger \hat a_j - \hat a_j^\dagger \hat a_k \right), \nn \\
& \hat Z_k= \hat a_k^\dagger \hat a_j + \hat a_j^\dagger \hat a_k,
\ee 
with $k=1,2,3$ and $j=k +1 \mod 3$, which form a generalized angular momentum algebra 
(cf. e.g. \cite{07phase,Nemo01}). 

A convenient choice of basis is given by the eigenstates of the Casimiroperator $4 \hat N(\hat N/3 +1)$
with $\hat N=\sum_{i=1}^3 \hat a_i^\dagger \hat a_i$ and the two elements of the Cartan subalgebra
$\hat X_1$ and $\hat X_2$ with
\be
\hat n_k \ket{n_1,n_2, n_3=N-n_1-n_2}=n_k \ket{n_1,n_2, n_3}.
\ee 
In this basis the reference state is chosen to be the
highest weight state  $\ket{N,0,0}$ and the generalized coherent 
states $\ket{\Omega}$ are obtained by the natural extension of 
(\ref{eqn-blochstate}) to higher dimensional rotations. 
Note, that $SU(3)$ is a two instead of one parameter Lie group, thus the parameter space 
of generalized the coherent states is now isomorphic to $SU(3)/U(2)$ and therefore four-dimensional.
In order to visualize the phase space dynamics we must hence use projections and Poincar\'e sections,
which are not necessarily be unique. 
For further mathematical details, see e.g. \cite{Nemo01} for SU(3), as well as \cite{Nemo00b,07phase} 
for the general case $SU(M)$.

However, for any actual calculation or numerical simulation the equivalent parametrization of
the generalized coherent states $\ket{\Omega}$ via the complex mode amplitudes $\psi_i$ 
\be
\ket{\Omega}=\frac 1 {\sqrt {N!}} \left(\sum_{i=1}^3 \psi_i \hat a_i^\dagger\right)^N \ket{0,0,0}
\ee
proves to be more practical. Using the conservation of the total particle number which corresponds to 
the normalization $\|\psi \|^2 = |\psi_1|^2+|\psi_2|^2+|\psi_3|^2 = 1$, we can fully characterize
the state vector by the occupations $p_{2} = |\psi_2|^2$, $p_3 = |\psi_3|^2$ and the relative phases
$q_1 = \arg(\psi_2)- \arg(\psi_1)$ and $q_3 = \arg(\psi_2)- \arg(\psi_3)$, as discussed in the case
of the two mode system. Therefore we will use this parametrization in the following.

Note, that such a description in terms of $SU(3)$ coherent states has already proven useful 
in the case of a circular arrangement with uniform tunneling elements $\Delta_{ij}=\Delta=const.$ \cite{Nemo00a}
and in the study of quantum discrete vortices \cite{Lee06}.
A comparison of these results to calculations based on common Glauber coherent states can be found in
\cite{Fran02}.

\section{Phase space distributions}
\label{sec-phasespace}

\begin{figure}[tb]
\centering
\includegraphics[width=8cm, angle=0]{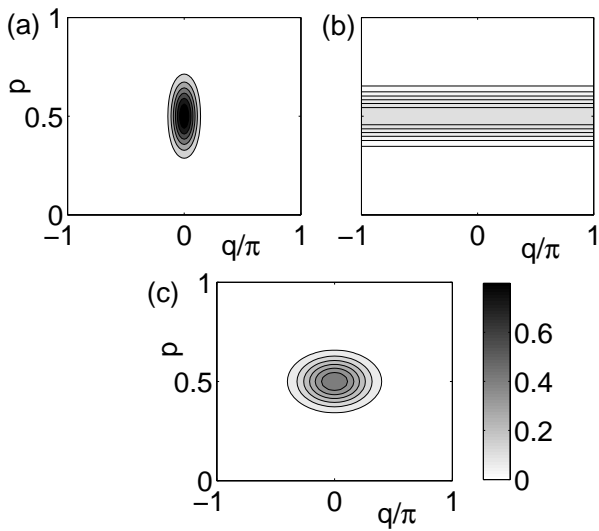}
\caption{\label{fig-husimi-examples}
Husimi functions in a Mercator projection of 
(a) the coherent state $\ket{p=1/2,q=0}$, 
(b) the Fock state $\ket{n_1 = N/2,n_2=N/2}$ and 
(c) the ground state of the Hamiltonian (\ref{eqn-hami-bh}) 
for $\Delta = 1$, $U=10$ and $\epsilon=0$. 
The number of particles is $N=40$ in all plots.}
\end{figure}

With the help of the generalized coherent states, which will be denoted $\ket{\Omega}$ 
in the following (independently of the corresponding dynamical group), one can readily 
introduce quasi phase space distributions for a BEC in a double or triple well trap. 
The Glauber-Sudarshan $P$-distribution is defined as the diagonal representation of 
the density operator $\hat \rho$ in generalized coherent states
\be
  \hat \rho = \int_X P(\Omega) \ket{\Omega}\bra{\Omega}  d\mu(\Omega),
\ee
where $\Omega$ represents the respective parametrization and $d\mu(\Omega)$ stands for the 
invariant measure on the respective phase space, denoted by $X$. Due to the overcompleteness 
of the coherent states, the $P$--function does always exist but is usually not unique. 
Furthermore it is not positive definite and often highly singular. On the other hand, the Husimi 
$Q$--function defined as the expectation value of the density operator in generalized coherent 
states,
\be
  Q(\Omega) = \bra{\Omega} \hat \rho \ket{\Omega},
\ee
is unique, regular and positive definite. Thus the $Q$-function
is especially suited for illustrations, while both quasi distribution functions 
will be used for actual calculations. However, the $Q$-function is also 
not a probability distribution function in the strict sense since it does not give the 
correct marginal distributions.
Note that it is also possible to define the Wigner function on a spherical phase space
but this leads to much more complicated expressions than the $P$- and $Q$-function,
since its construction uses harmonic functions on the respective phase space. 
This makes actual calculations a hard task even for two lattice 
sites corresponding to $SU(2)$ (see, e.g., the contradictory results in \cite{Klim02} 
and \cite{Zuec07}) and almost impossible for larger systems.
Furthermore, a positive $P$-representation respecting $SU(M)$ symmetry has 
been introduced and analyzed in \cite{Barr08}.

To get a first impression of the phase space distributions considered here 
we have plotted the Husimi function of the two mode system for (a) the coherent 
state $\ket{p=1/2,q=0}$, (b) the Fock state $\ket{n_1 = N/2,n_2=N/2}$ and 
(c) the ground state of the Hamiltonian (\ref{eqn-hami-bh}) for 
$\Delta = 1$ and $U=10$ in a Mercator projection of the sphere in 
Fig.~\ref{fig-husimi-examples}.
The coherent state is maximally localized at the position $(p,q)$ and
thus closest to a point in classical phase space.
On the contrary a Fock state is localized around $p = n_2/N$. The phase $q$ 
is completely delocalized in the sense that the Husimi function is uniform in
$q$. 
Finally one can clearly visualize the number squeezing of the eigenstate (c) 
in comparison with the coherent state (a) due to the interaction term in the 
Hamiltonian -- a fact which is desirable for matter wave interferometer 
experiments \cite{GBJo07}.

\begin{figure}[tb]
\centering
\includegraphics[width=8cm, angle=0]{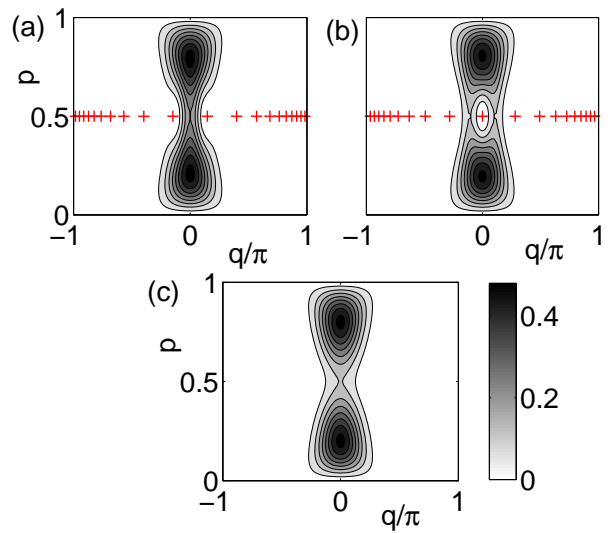}
\caption{\label{fig-catstate}
(Color online) 
Husimi density of the cat states $\ket{\psi_\pm} \propto \ket{0.2,0} \pm \ket{0.8,0}$
(a),(b) and the incoherent sum $\hat \rho_+= (\ket{0.2,0} \bra{0.2,0} + \ket{0.8,0} \bra{0.8,0})/2$
(c) for $N=20$ particles. The zeros of the Husimi function are marked by red crosses.}
\end{figure}

The overcompleteness of the basis of (generalized) coherent states guarantees that
one can uniquely reconstruct the quantum state out of the quasi probability 
distribution. It is even possible to determine a pure state solely from the 
$N$ zeros of the Husimi function or the Bargmann function up to a global phase 
factor \cite{Gnut01}. Thus these zeros carry the essential information about 
the quantum state and its coherence properties. 
As an example, Fig.~\ref{fig-catstate} compares the Husimi distribution of 
a cat state in the two mode case, i.e.~a coherent superposition of two $SU(2)$ 
coherent states $\ket{\psi_\pm} \propto \ket{p = 0.2, q=0} \pm \ket{p = 0.8,q = 0}$,
with the incoherent sum of these two states 
$\hat \rho_+ = (\ket{0.2,0} \bra{0.2,0} + \ket{0.8,0} \bra{0.8,0})/2$.
The global shape of the Husimi function of the three states 
appears quite similar with significant differences only around
the point $(p,q) = (0.5,0)$. The Husimi density is increased
for the coherent superpositions $\ket{\psi_+}$ in (a) and 
decreased for $\ket{\psi_-}$ in (b) in 
comparison with the incoherent sum, reflecting constructive or destructive 
interference, respectively. The complete information about the pure states
$\ket{\psi_\pm}$ is coded in the distribution of the $N=20$ zeros of the 
Husimi function, which are plotted as red crosses in the figure.
The Husimi function of the mixed state $\hat \rho_+$ has no zero at all,
whereas the Husimi function of a coherent state (compare also 
Fig. \ref{fig-husimi-examples}) has a single $N$-fold degenerate zero
opposite to the maximum of the quasi probability distribution.

\section{Dynamics and the macroscopic limit}
\label{sec-dynamics}

Evolution equations for the classical phase space distributions can be derived
using the mapping of operators in Hilbert space to differential operators on the
classical phase space. The $\Dop^\ell$-algebra representation of an arbitrary 
hermitian operator $\hat A$ is defined such that the resulting differential operator 
acting on the parametrization of the generalized coherent state projector has exactly 
the same effect as the original operator:
\be
  \hat A \ket{\Omega} \bra{\Omega} & \stackrel{!}{=} &\Dop^\ell(\hat A) \ket{\Omega} \bra{\Omega}, \\
  \ket{\Omega} \bra{\Omega} \hat A & \stackrel{!}{=} &\Dop^\ell(\hat A)^* \ket{\Omega} \bra{\Omega}.
\ee 
Here, the superscript $\ell$ refers to side (left) from which the operator acts. 
The $\Dop^\ell$-algebra representation is well known for the Heisenberg-Weyl algebra.
It has been introduced for the $su(M)$-algebra by Gilmore and coworkers (see \cite{Zhan90}
and references therein). Here, we just state the expression for the generators of the 
$su(2)$-algebra in the angular parametrization (\ref{eqn-blochstate}):
\be \label{eqn_dellsu2}
  \Dop^\ell(\hat L_+) & = & e^{\ri \phi} \left( \frac N 2 \sin \theta
     + \frac \ri 2 \cot \frac \theta 2 \abl{\phi} 
     +  \cos^2 \frac \theta 2 \abl{\theta}  \right), \nn \\
  \Dop^\ell(\hat L_-) & = & e^{-\ri \phi} \left( \frac N 2 \sin \theta
     - \frac \ri 2 \tan \frac \theta 2 \abl{\phi} 
     - \sin^2 \frac \theta 2 \abl{\theta}   \right), \nn \\
  \Dop^\ell(\hat L_z) & = & - \frac{N}{2} \cos \theta
     + \frac{1}{2} \sin \theta  \frac{\partial}{\partial \theta} 
     + \frac \ri 2 \frac{\partial}{\partial \phi} \, .
\ee

Independently of the number of modes $M$, the evolution equation for the $Q$--function 
can be expressed in terms of the differential operators $\Dop^\ell$:
\be \label{eqn_evoldopQ}
  \frac{\partial}{\partial t} Q(\Omega) &=& \tr( \dot{\hat \rho} \ket{\Omega}\bra{\Omega} ) \nn \\
   &=& -i \, \tr ( \hat \rho \hat H \ket{\Omega}\bra{\Omega} 
          - \ket{\Omega}\bra{\Omega} \hat H \hat \rho) \\
   &=& -i (\Dop^\ell(\hat H) - \Dop^\ell(\hat H)^*) Q(\Omega). \nn 
\ee
For the $P$-function we need an associated $\tilde \Dop^\ell$-algebra defined by
an integration by parts:
\be
  && \int_{X} \{P(\Omega) \Dop^\ell(H) \}\ket{\Omega}\bra{\Omega} d\mu(\Omega) = \nn \\
  && \qquad \qquad  \int_{X} \{\tilde \Dop^\ell(H) P(\Omega) \}\ket{\Omega}\bra{\Omega} d\mu(\Omega).
\ee
Using this algebra, the evolution equation for the $P$-function can be derived via
\be \label{eqn_evoldopP}
  \dot{\hat \rho} &=& \int_{X} \frac{\partial}{\partial t}  P(\Omega) \ket{\Omega}\bra{\Omega} d\mu(\Omega) = \\
  &=& \int_{X} i \left( \tilde \Dop^\ell(H) - \tilde \Dop^\ell(H)^*\right) 
          P(\Omega) \ket{\Omega}\bra{\Omega} d\mu(\Omega). \nn
\ee

\begin{figure}[tb]
\centering
\includegraphics[width=8cm, angle=0]{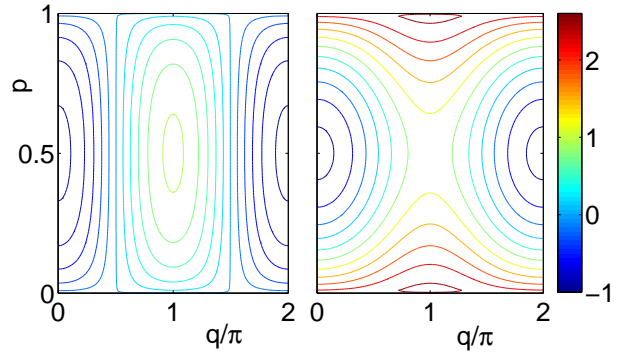}
\caption{\label{fig-phasespace}
(Color online)
Classical phase space structure: Lines of constant energy $\HH(p,q)$
for $\Delta = 1$, $\epsilon = 0$, and $g = 0$ (left) resp. $g = 10$ (right).}
\end{figure}

The $\Dop^\ell$-algebra representation of the relevant operators and the evolution 
equation for the $M$-site Bose-Hubbard-Hamiltonian have been calculated in the preceding 
paper \cite{07phase}. They are generally applicable to every problem within this symmetry class.

\subsection{Mean-field dynamics of the two-mode system}

Since the mapping to the differential operators is linear, the explicit evolution equations for the $Q$- and $P$-function can be directly calculated from the imaginary part (cf. Eqn. (\ref{eqn_evoldopQ}) and (\ref{eqn_evoldopP})) of the $\Dop^\ell$ operators (see Eqn. (\ref{eqn_dellsu2})\,). This yields the 
following results:
\be \label{eqn_qqevolutionexact}
  && \frac{\partial}{\partial t}  Q(p,q) = \bigg\{ + 2 \epsilon \frac{\partial}{\partial q}  
        \label{eqn-eom-husimi}   \\
   && \qquad   + \Delta \left( 2\sqrt{p-p^2} \sin q \frac{\partial}{\partial p}
        + \frac{1-2p}{\sqrt{p-p^2}} \cos q \frac{\partial}{\partial q} \right) \nn \\
    && \qquad + U \left( N(1-2p) - 2p(1-p) \frac{\partial}{\partial p} \right) 
          \frac{\partial}{\partial q} \bigg\} Q(p,q) \nn
\ee
for the Husimi $Q$-function and 
\be
  && \frac{\partial}{\partial t}  P(p,q) = \bigg\{ + 2 \epsilon \frac{\partial}{\partial q}  
        \label{eqn-eom-p}   \\
   && \quad + \Delta \left( 2\sqrt{p-p^2} \sin q \frac{\partial}{\partial p}
        + \frac{1-2p}{\sqrt{p-p^2}} \cos q \frac{\partial}{\partial q} \right) \nn \\
    && \quad + U \left( (N+2)(1-2p) + 2p(1-p) \frac{\partial}{\partial p} \right) 
          \frac{\partial}{\partial q} \bigg\} P(p,q) \nn
\ee
for the Glauber-Sudarshan $P$-function.

One can directly show that both evolution equations conserve the normalization. 
Furthermore it can be shown that the interaction terms $\sim U$ cancel exactly 
for $N=1$, since a single Boson obviously does not interact.
The expectation values 
of the angular momentum operators (\ref{eqn-angular-op}) are exactly given
by the phase space averages \cite{07phase}
\be
  && \langle \hat J_k \rangle = (N+2) \int_{\Szwo} s_k(p,q) Q(p,q) d\mu(p,q) 
   \quad \mbox{and}  \nn \\
   && \langle \hat J_k \rangle = N \int_{\Szwo} s_k(p,q) P(p,q) d\mu(p,q),
  \label{eqn-pq-exvalues}
\ee
where $k = x,y,z$ and
\be
  \vec s = 
  \left( \begin{array}{c}
    \sqrt{p(1-p)} \cos(q) \\ \sqrt{p(1-p)} \sin(q) \\ p-1/2
  \end{array} \right)
\ee
is the classical Bloch vector.

\begin{figure}[tb]
\centering
\includegraphics[width=8cm, angle=0]{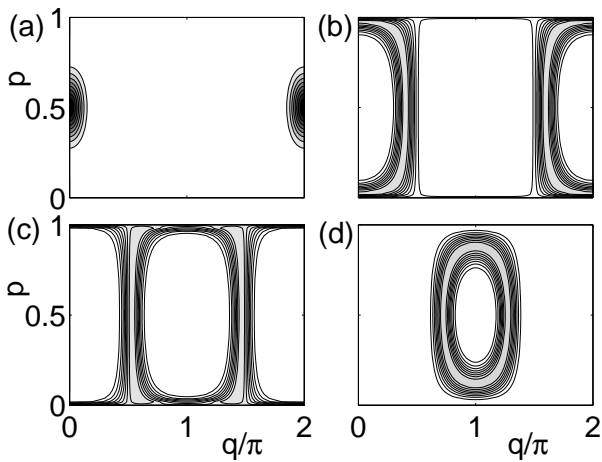}
\caption{\label{fig-eigenstate-g0}
Husimi density of the eigenstates $\ket{E_n}$ with $n = 1,15,23,34$ (a-d)
of the Hamiltonian (\ref{eqn-hami-bh}) for $\Delta = 1$, $\epsilon = 0$,
$UN = 0$ and $N=40$ particles. Dark colors encode high values of the 
Husimi density.}
\end{figure}

These exact equations provide a valuable starting point to study the limit of large 
particle numbers, since it emerges quite naturally in the phase space picture.
In the macroscopic limit $N \rightarrow \infty$ with $g = UN$ fixed, the second 
order derivative terms in the evolution equations become negligible, as they 
are vanishing as $\mathcal{O}(1/N)$.
The appearance of a non-positive diffusion term as the difference between the 
exact quantum dynamics and the classical dynamics has already been discussed 
in \cite{Milb02}. Thus, in the macroscopic case one obtains a Liouville 
equation for a quasi-classical phase space distribution function $\rho(p,q)$: 
\be \label{eqn_liouvsu2}
  \frac{\partial}{\partial t} \rho(p,q) &=& - \left( \frac{\partial \HH}{\partial p} \frac{\partial}{\partial q}
     - \frac{\partial \HH}{\partial q} \frac{\partial}{\partial p} \right) \rho(p,q) \nn \\
&=& - \{\HH(p,q),\rho(p,q)\} \label{eqn-liouville}
\ee
with the Poisson bracket $\{ \cdot,\cdot \}$ and the Gross-Pitaevskii Hamiltonian function
\be
   \HH = -2 \epsilon p - 2 \Delta \sqrt{p(1-p)} \cos(q) + \frac{g}{4} (1-2p)^2.
   \label{eqn-ham-clas}
\ee
The macroscopic interaction strength depends upon the operator ordering and is thus
given by $g=UN$ if we start from the $Q$-function and by $\tilde g = U(N+2)$ for the 
$P$- function. This difference also vanishes in the macroscopic limit 
$N \rightarrow \infty$ with $g = UN$ fixed.
Therefore the phase space picture gives rise to a Liouvillian flow for the time-evolution of the quasi 
probability distribution in the macroscopic limit.

\begin{figure}[tb]
\centering
\includegraphics[width=8cm, angle=0]{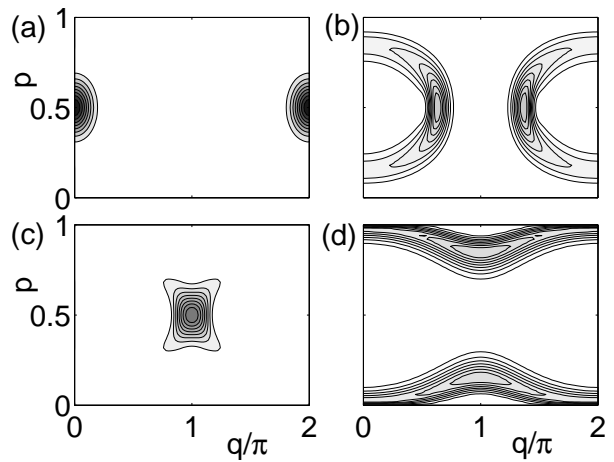}
\caption{\label{fig-eigenstate-g10}
As Fig.~\ref{fig-eigenstate-g0}, however for $UN = 10$.}
\end{figure}

The common mean-field limit is subject to even more severe restrictions. To circumstantiate
this point, we will have a closer look at the relation between the Liouville dynamics and the GPE.
To obtain the latter, one must not only take the macroscopic limit of the evolution equations 
(\ref{eqn-eom-husimi}) and (\ref{eqn-eom-p}), but also assume a pure condensate, respectively a 
$SU(M)$ coherent state which is maximally localized during the whole time evolution. 
In this case the dynamics of a phase the space distribution of the initial state is approximated by a 
point distribution, respective a single trajectory instead of an extended distribution.
Therefore, any information about quantities involving higher moments of the state is lost.

So, if we assume a delta-distribution $\rho(p',q')=\delta(p'-p)\delta(q'-q)$, 
we can derive the GPE starting from the Liouville equation (\ref{eqn_liouvsu2}). 
This yields the canonical equations of motion
\be \label{eqn-eom-ham}
  && \dot p =- \frac{\partial \HH}{\partial q} = - 2 \Delta \sqrt{p(1-p)} \sin(q) \\
  && \dot q = \frac{\partial \HH}{\partial p} = 
      - 2 \epsilon - \Delta \frac{1-2p}{\sqrt{p(1-p)}} \cos(q) - g (1-2p). \nn
\ee
These equations are equivalent to the celebrated discrete Gross-Pitaevskii equation
\be
  i \frac{d}{dt} \left(\begin{array}{c} \psi_1 \\ \psi_2 \end{array} \right)
  = \left(\begin{array}{c c} \epsilon + g |\psi_2|^2 & -\Delta \\ 
             -\Delta & -\epsilon + g |\psi_2|^2 \end{array} \right)
  \left(\begin{array}{c} \psi_1 \\ \psi_2 \end{array} \right),
  \label{eqn-gpe}
\ee
if one identifies $\psi_1 = \sqrt{1-p} $ and $\psi_2 = \sqrt{p} \exp{(-iq)}$ and
neglects the global phase. 
In the non-interacting case $U=0$, these mean-field equations of motion are exact
and an initially coherent state remains coherent in time. Hence, no information is
lost, since the higher moments can be easily reconstructed.
However, in the interacting 
case an initially pure BEC typically deviates quite rapidly from the set of maximally 
localized states. Moreover, by approximating the state by a product state one looses the entire
information about the higher moments of the state. Therefore many quantities of interest as
for example variances etc.~are not accessible, whereas they can be easily estimated using 
the Liouville dynamics and phase space averages similar to Eqn. (\ref{eqn-pq-exvalues}).

The classical phase space structure induced by the Hamiltonian function 
(\ref{eqn-ham-clas}) is shown in Fig.~\ref{fig-phasespace} for 
$\Delta=1$, $\epsilon = 0$ and two different value of the interaction strength, 
$g = 0$ and $g = 10$. In the linear case one recovers 
the simple Rabi or Josephson oscillations. One of the elliptic fixed points 
bifurcates if the interaction strength exceeds the critical value $g = 2\Delta$, 
leading to the self-trapping effect \cite{Milb97,Smer97,Ragh99}.

Furthermore, we have plotted the  Husimi function for some selected
eigenstates of the Bose-Hubbard Hamiltonian (\ref{eqn-hami-bh}) for the 
linear case $UN = 0$ in Fig.~\ref{fig-eigenstate-g0} and for $UN = 10$ 
in Fig.~\ref{fig-eigenstate-g10}. The Husimi functions of the eigenstates 
localize on the classical phase space trajectories of the respective 
energy as shown in Fig.~\ref{fig-phasespace}, where their sum covers the 
classical phase space uniformly, 
\be
  \sum_n Q_{\ket{n}}(p,q) = 1.
\ee
In the linear case $UN = 0$, the classical trajectories correspond 
to the Josephson oscillations between the two wells as shown on the 
left-hand side of Fig.~\ref{fig-phasespace}. This behavior is directly 
mirrored by the eigenstates as shown in Fig.~\ref{fig-eigenstate-g0}:
The lowest eigenstate localizes on the classical fixed point $(p,q) = (0,0)$ 
while the others correspond to Josephson oscillations. 

The situation is more interesting, when the interaction strength exceeds 
the critical value for the self-trapping bifurcation, as shown exemplarily
for $UN = 10$ in Fig.~\ref{fig-eigenstate-g10}. For low energies one still 
finds Rabi modes, while the eigenstates of higher energies correspond
to self-trapping trajectories with a persistent particle number difference
and a running phase (cf.~Fig.~\ref{fig-phasespace}, right). However, the 
eigenstates always localize equally on the trajectories with $p > 1/2$ and 
$p<1/2$ because of the symmetry of the Hamiltonian (\ref{eqn-hamiltonian-2level}). 
Most interestingly, quantum states can also localize at the unstable hyperbolic 
fixed point as shown in Fig.~\ref{fig-eigenstate-g10} (c).

\subsection{Mean-field dynamics of the three-mode system}

As in the case of the two-mode system, the exact evolution equations of the phase space
distribution functions can be calculated from the $\Dop^\ell$ operators.
The results given in \cite{07phase} are considerably more complicated
as in the case of only two modes. 
However, in the macroscopic limit one again recovers a Liouville equation 
\be
  \frac{\partial}{\partial t} \rho(p_1,p_3,q_1,q_3) &=& - \{\HH,\rho(p_1,p_3,q_1,q_3)\}
\ee
with the macroscopic Hamiltonian function
\begin{eqnarray}
  {\cal H} & = & \epsilon_1 p_1 + \epsilon_3 p_3
   + \frac{g}{2} \left (p_1^2 + p_3^2 + (1-p_1-p_3)^2 \right) \nn \\
   && -2 \Delta_{12} \sqrt{1-p_1-p_3} \sqrt{p_1} \cos(q_1) 
   \label{eqn-classical-hamiltonian} \\
   && -2 \Delta_{23} \sqrt{1-p_1-p_3} \sqrt{p_3} \cos(q_3). \nn 
\end{eqnarray}

\begin{figure}[bt]
\centering
\subfigure[$g=-5$ and ${\cal H} = 1$]{
\includegraphics[width=7cm,  angle=0]{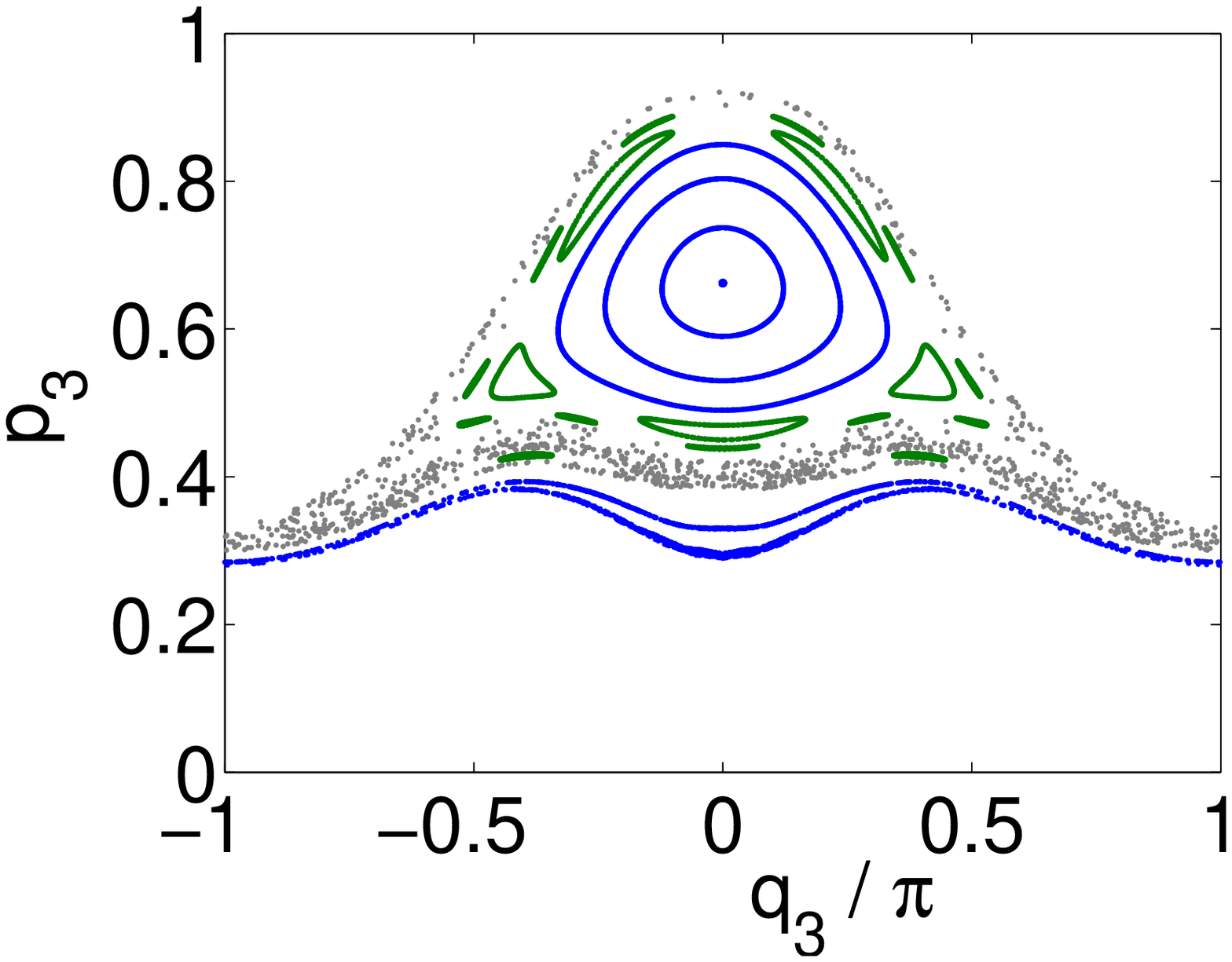}}
\subfigure[$g=-10$ and ${\cal H} = -1$]{
\includegraphics[width=7cm,  angle=0]{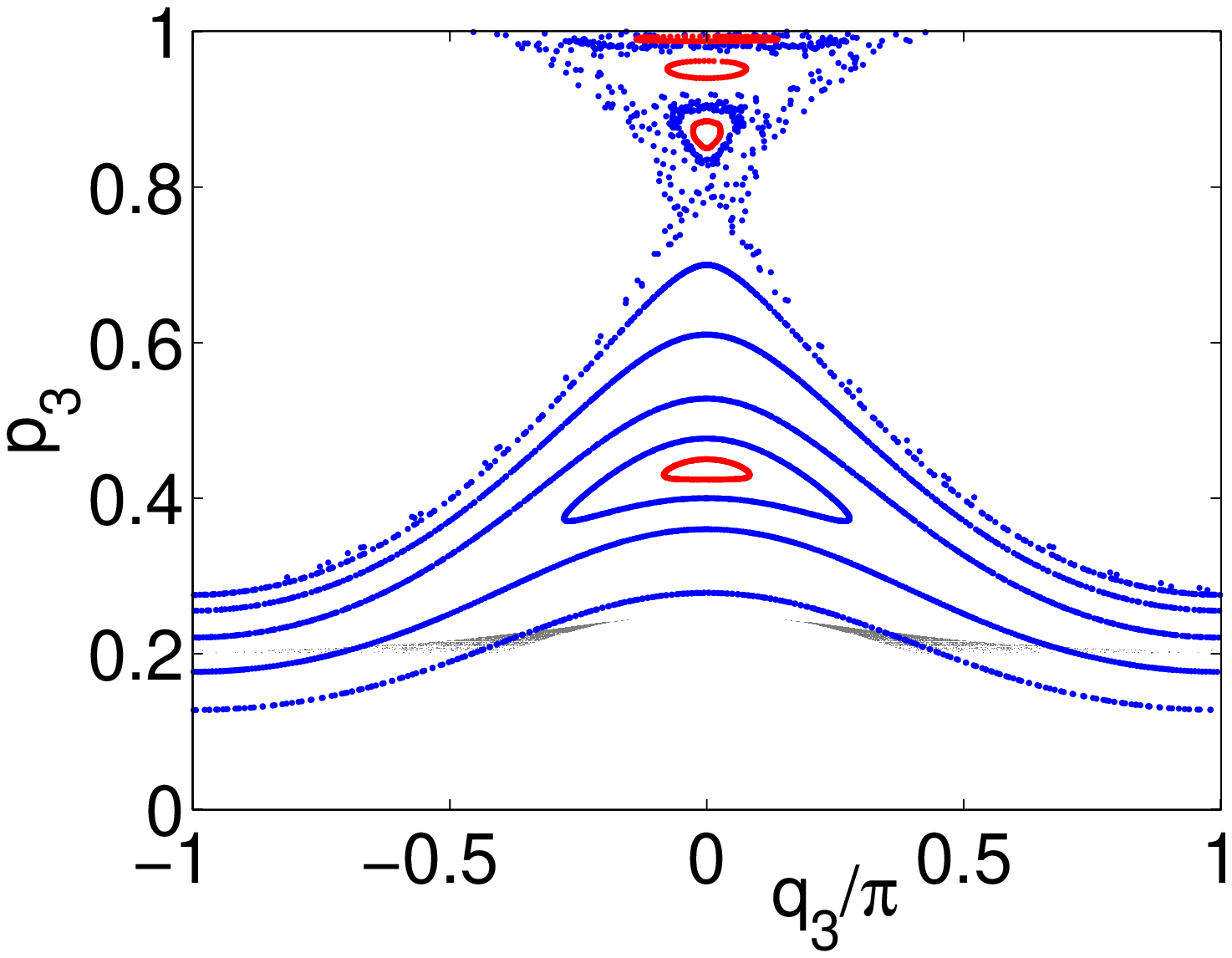}}
\caption{\label{fig-pphase-g5} \label{fig-pphase-g10}
(Color online) Poincar\'e section of the classical phase space ($q_1 = 0$,
see text for details) for $\epsilon_1 = 2$, $\epsilon_3 = 4$, 
$\Delta_{12} = \Delta_{23} = 1$, $g=-5$ and a fixed total energy ${\cal H} = +1$ 
(a) and $-g=10$ and ${\cal H} = -1$ (b). Trajectories around the elliptic 
fixed points are coloured in red to guide the eye.}
\end{figure}

The main new feature of the dynamics in three-levels systems is that the mean-field 
dynamics can become classically chaotic (cf. e.g. \cite{Buon03,Fran03,05level3,Moss06} 
and references therein), where the transition goes from regular dynamics for $g=0$, to
mixed chaotic dynamics for stronger nonlinearities $|g|$, whereas the 
mean-field dynamics of the two mode system with constant parameters is always regular. 
Figure \ref{fig-pphase-g5} (a) shows a Poincar\'e section of the corresponding
phase space for $g=-5$ and ${\cal H} = 1$, where the dynamics is still
mostly regular. However, a classically chaotic strip is found between
the elliptic fixed point at $(p_3,q_3) = (0.66,0)$ and the running phase
modes below. Chains of large regular islands are found within this chaotic sea
(colored in green in the figure). 
Another example is shown in Fig.~\ref{fig-pphase-g5} (b) for $g=-10$ and 
${\cal H} = -1$. Again we observe elliptic fixed points with large values of
$p_3$ separated from the running phase modes by a chaotic region in phase 
space. However, the regular running phase modes now occupy most of the 
energetically allowed region of phase space. Furthermore one observes a 
novel fixed point embedded in this region of phase space.

This phase space structure is typical for the route to chaos in Hamiltonian 
systems described by the KAM and the Poincar\'e-Birkhoff theorem.
The dynamics becomes chaotic in the phase space region close to the
separatrix. Resonant tori are destroyed and decompose into a series
of elliptic and hyperbolic fixed points, where the elliptic ones form the
island chains embedded in the chaotic region. The remaining invariant tori
confine this chaotic sea. 

Note that the use of a Poincar\'e section is not straightforward
for the dynamical system considered here. As usual, the variables
$(p_3,q_3)$ are plotted when the trajectories intersect the
$q_1 = 0$-plane with $\dot q_1 > 0$.
The remaining dynamical variable $p_1$ follows from energy conservation.
However the uniqueness has to be checked explicitly. For example, the 
assignment of $p_1$ is not unique in Fig.~\ref{fig-pphase-g5} (b) 
in a small area of phase space which is shaded in grey in the figure. 
Solving equation (\ref{eqn-classical-hamiltonian}) for $p_1$ gives two 
possible solutions. In order to make the calculation of $p_1$ unique we 
define that only the solution with smaller value of $p_1$ is plotted in the Poincar\'e section.
Furthermore, large regions of the $(p_3,q_3)$-plane are energetically
forbidden for $q_1 = 0$ (cf. Fig.~\ref{fig-pphase-g5}).

\begin{figure}[bt]
\centering
\includegraphics[width=7cm,  angle=0]{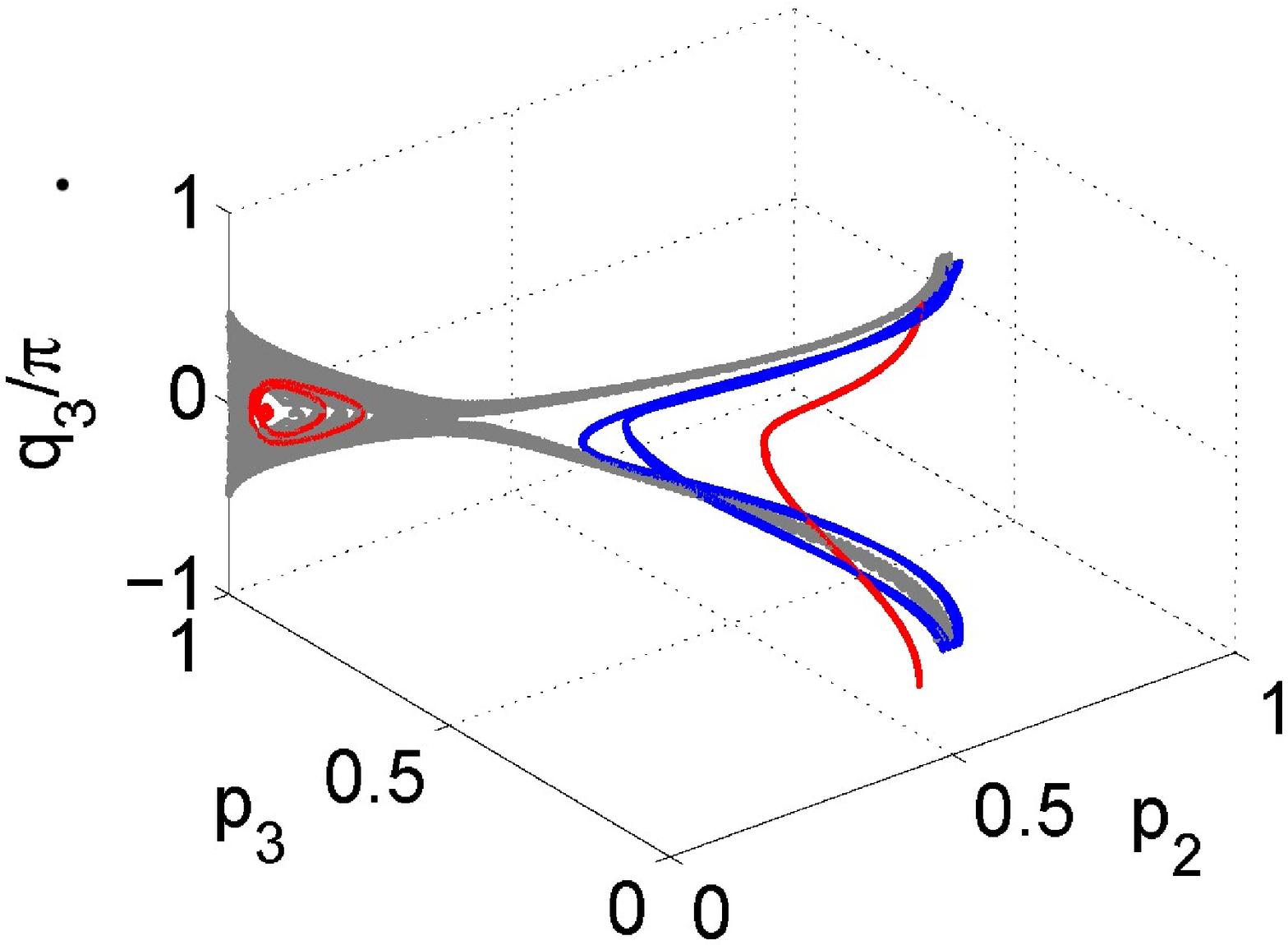}
\includegraphics[width=7cm,  angle=0]{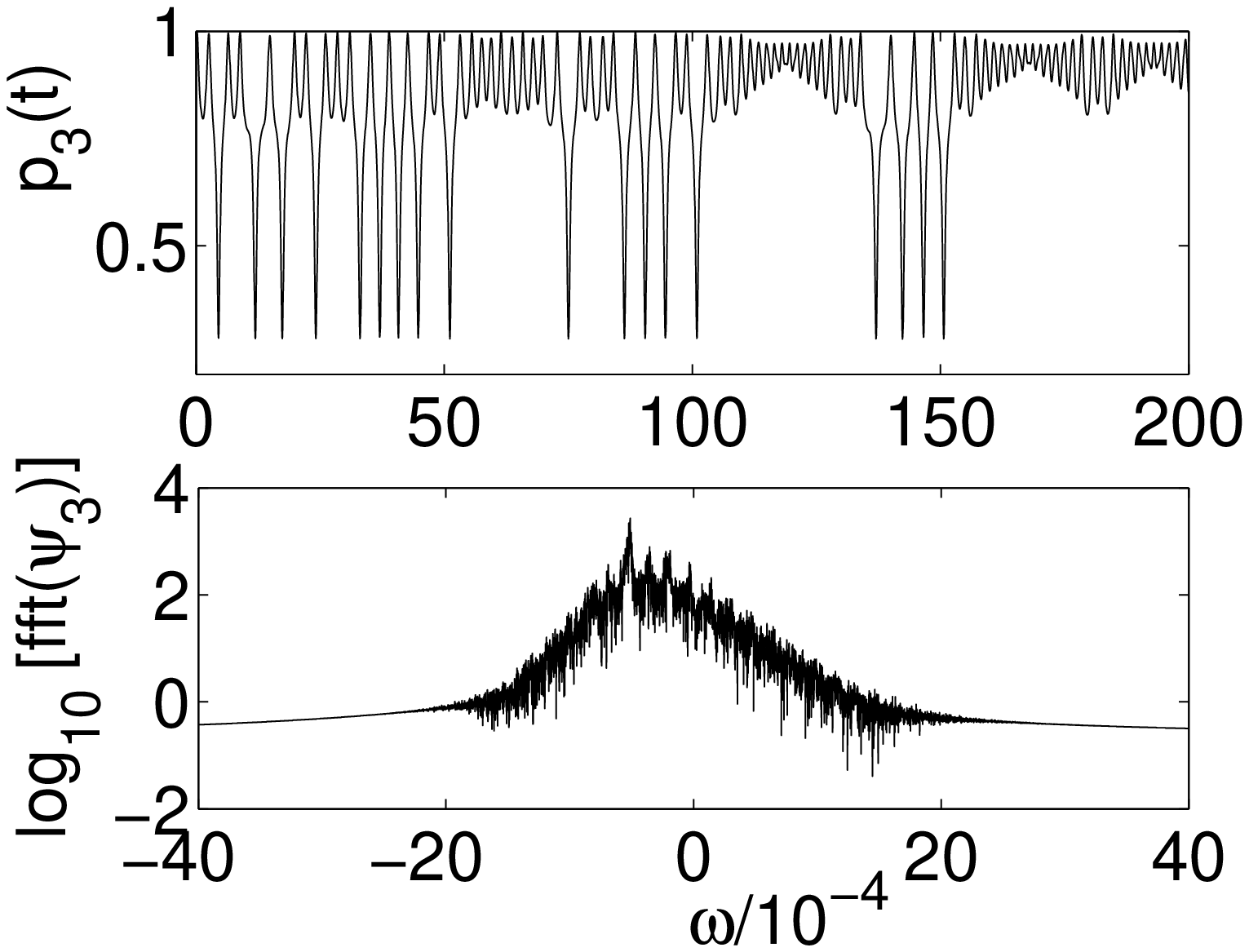}
\caption{\label{fig-3phase-g10} \label{fig-dyn-g10}
(Color online) 
Upper panel: mean-field phase space trajectories for $\epsilon_1 = 2$, 
$\epsilon_3 = 4$, $\Delta_{12} = \Delta_{23} = 1$, $g=-10$ and a fixed 
total energy ${\cal H} = -1$.
Lower panel: Time series ($p_3(t) = |\psi_3(t)|^2$) and power spectrum
of a trajectory in the chaotic sea.}
\end{figure}

Let us have a closer look at the mixed classically-chaotic phase space
shown in Fig.~\ref{fig-pphase-g10} (b). The dynamics is also shown in 
a three-dimensional phase space plot in Fig.~\ref{fig-3phase-g10}.
Only few trajectories are plotted exemplarily. The periodic orbits
(coloured in red) correspond to the elliptic fixed points in
the Poincar\'e section (Fig.~\ref{fig-pphase-g10}).
Two periodic orbits are found on the left-hand side of the figure,
oscillating with a small amplitude both in the populations and in
the relative phases.
Another periodic trajectory shows a larger amplitude of the oscillations
of the populations $p_2$ and $p_3$ and a running phase $q_3$. This periodic
trajectory is embedded in a set of similar quasi-periodic ones,
of whom one is plotted as a blue line in the figure.
Furthermore, a trajectory in the chaotic sea is plotted in grey. It is
observed that the chaotic sea splits into two distinct regions with
narrow links: One around the periodic orbits on the left and one in the
vicinity of the running-phase trajectory on the right-hand side.

\begin{figure*}[tb]
\centering
\includegraphics[width=16cm, angle=0]{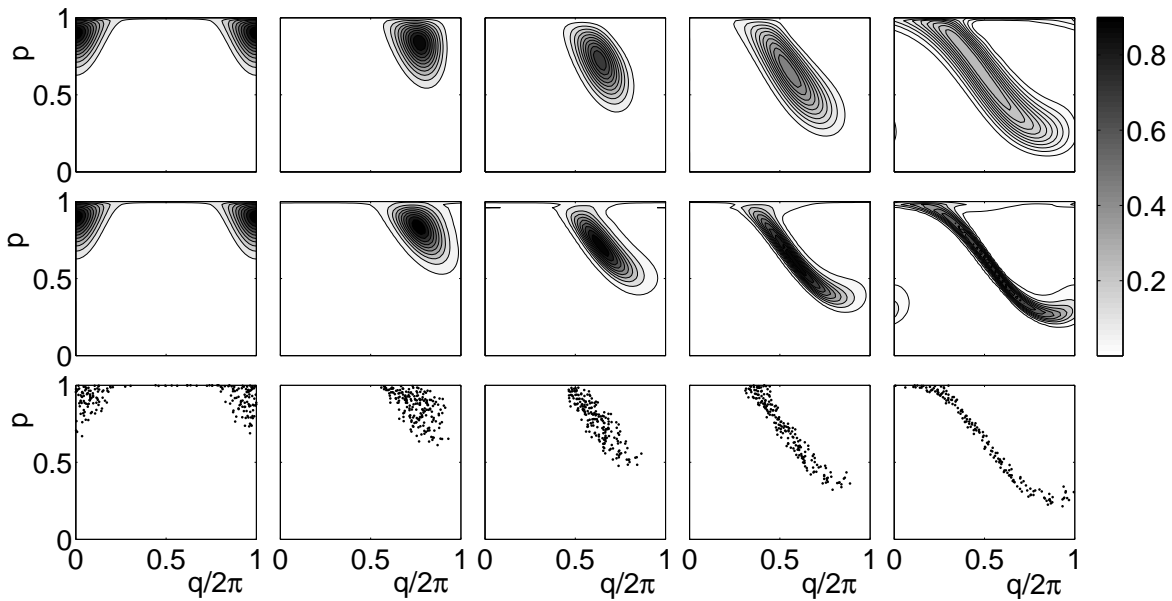}
\caption{\label{fig-zusammenbruchps}
Breakdown of the mean-field approximation around a hyperbolic fixed point
in phase space. Shown is the dynamics of an initially coherent state located 
at $p_0 = 0.9045$ and $q_0 = 0$ at times $t = 0,0.12,0.24,0.36,0.48$ (from 
left to right) for $\Delta=1$, $UN = 10$ and $N = 20$ particles.
The exact quantum evolution of the Husimi function (top) is compared to the
classical Liouvillian dynamics (middle) and to the dynamics of a classical
phase space ensemble (bottom).}
\end{figure*}

This 'bottleneck' effect leads to an intermittent dynamics: A trajectory
can be confined to certain regions in the chaotic sea for a long time,
where its time evolution looks quasi regular. Chaotic bursts to the
remaining part of the chaotic sea are found between the quasi-regular
oscillations. This is shown in Fig.~\ref{fig-dyn-g10}, where the time
series $p_3(t)$ for one particular trajectory in the chaotic sea is plotted
as well as the corresponding power spectrum. One observes long
quasi-regular oscillations with a small amplitude intermittent
by large amplitude bursts. Physically this means that the BEC mostly stays
in the right well for a long time with irregular bursts to the middle well.

In conclusion, we stress that in any case, two steps of approximation were necessary to 
derive the Gross-Pitaevskii equation: In the first step we have neglected the quantum 
noise term in the evolution equations (\ref{eqn-eom-husimi}) and (\ref{eqn-eom-p})
which is always possible in the macroscopic limit $N \rightarrow \infty$ with
$g \equiv UN$ fixed. Only in the second step to derive the GPE one assumes a nearly pure 
condensate, so that its phase space representation is strongly localized and can 
be approximated by a point distribution or a single trajectory in phase space.

Thus it is not only possible to simulate the dynamics of a strongly localized product state,
but also of every other possible initial state using the Liouville equation while the error 
vanishes as $1/N$. Whereas the single-trajectory mean-field approximation will break down if 
the quantum state differs significantly from a product state, the phase space description still 
gives excellent results. This issue and its implications will be further investigated in the following.

\section{Analysis of dynamical instabilities and chaos}
\label{sec-breakdown}

The phase-space description of the Bose-Hubbard model is especially suited 
to explore the correspondence of the quasi-classical mean-field approximation 
and the many-particle quantum dynamics. The established derivations of
the Gross-Pitaevskii equation assume a fully coherent quantum state, so
that the mean-field approximation is restricted to this class of quantum
states. In fact it has been shown that the approximation is no longer valid
if the Gross-Pitaevskii dynamics becomes classically unstable 
\cite{Cast97,Cast98}.
In this section, we investigate the behaviour of the mean-field approximation
around localized, as well as extended regions of dynamical instabilities. 
It is shown, that the Liouville dynamics not only resolves well-known problems of the 
common mean-field description, but enables us to even study highly chaotic systems.

\subsection{Resolution of the break-down of mean-field}

\begin{figure}[tb]
\centering
\includegraphics[width=8cm, angle=0]{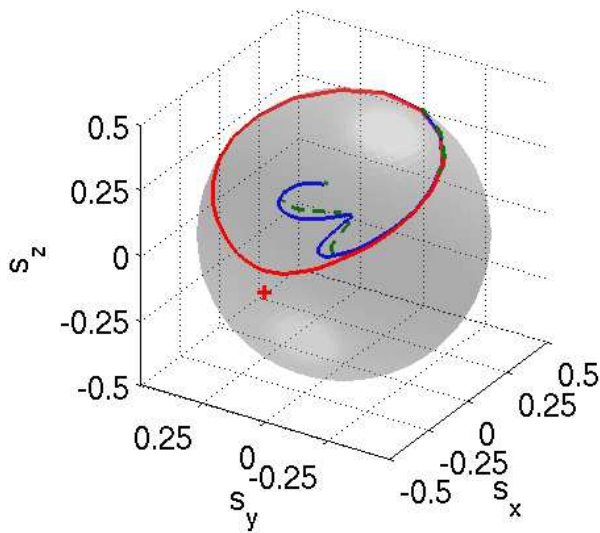}
\caption{\label{fig-breakdown}
(Color online) Time evolution of the Bloch vector $\langle \hat{\vec J} \rangle/N$ 
approaching the hyperbolic fixed point (solid blue line) for the same parameters as in 
Fig.~\ref{fig-zusammenbruchps}. The ensemble average over $500$ trajectories 
(dashed green line) closely follows the quantum result, while a single trajectory 
(solid red line) breaks away in the vicinity of the hyperbolic fixed point (marked by a 
cross).}
\end{figure}

A particular illustrative example of the failure of the common mean-field equations 
was introduced by Anglin and Vardi \cite{Vard01b,Angl01}, who demonstrated the breakdown 
of the mean-field approximation for a two-mode BEC around the hyperbolic
fixed point shown in Fig.~\ref{fig-phasespace} on the right-hand side.
Therefore, this is an excellent example to demonstrate that the approximation
by the Liouville dynamics clearly goes beyond the usual mean-field description 
and to enlight the reason of the break-down. 
The top row of Fig.~\ref{fig-zusammenbruchps} shows the resulting evolution 
of the quantum Husimi $Q$-function. Initially the system is in a fully 
condensed state, i.e. a $SU(2)$ coherent state at $p_0 = 0.9045$ and 
$q_0 = 0$, and thus the Husimi function is maximally localized. In the 
course of time the condensate approaches the hyperbolic fixed point, where 
the Husimi function rapidly diffuses along the unstable classical manifold. 
The coherence is lost and the Gross-Pitaevskii equation is no longer valid.
This is further illustrated in Fig.~\ref{fig-breakdown}, where the evolution
of the quantum Bloch vector $\langle \hat{\vec J}\rangle/N$ (solid blue line) 
is compared to its classical counterpart $\vec s$ (solid red line). Both 
agree well in the beginning, when the system is fully condensed. However,
the mean-field approximation breaks down as they approach the hyperbolic 
fixed point (marked by a cross in the figure). The quantum Bloch vector 
penetrates into the Bloch sphere, while the classical vector $\vec s$ is 
restricted to the surface.

However, this breakdown of the mean-field approximation is only due to the
neglect of higher moments of the quantum state and not a consequence of a failure
of the quasi-classical Gross-Pitaevskii dynamics. Thus it is easily resolved 
in quantum phase space. The middle row of Fig.~\ref{fig-zusammenbruchps} 
shows the evolution of a classical phase space distribution propagated by 
the Liouville equation (\ref{eqn-liouville}) which coincides with the Husimi 
function at $t=0$. With increasing particle number $N$ the width of the initial 
distribution decreases such that one recovers a single phase space point in the
macroscopic limit $N\rightarrow \infty$. One observes that the classical distribution 
captures the essential features of the quantum evolution -- the spreading along
the unstable manifold of the hyperbolic fixed point. Due to the neglect of
the second order differential term in Eqn.~(\ref{eqn_qqevolutionexact}) this spreading is
a little overestimated while the quantum spreading in the orthogonal direction 
is absent. Thus, this can be construed as a disregard of quantum noise which 
guarantees the uncertainty relation. 
Furthermore we can interpret the Liouvillian flow in terms of a classical phase 
space ensemble as shown in the bottom row of Fig.~\ref{fig-zusammenbruchps}.
At $t=0$ this ensemble is generated by 500 phase space points to mimic the
quantum Husimi distribution. Afterwards all trajectories evolve according to 
the Gross-Pitaevskii equation (\ref{eqn-gpe}) resp.~(\ref{eqn-eom-ham}). 
This approximation is comparable to numerical Monte Carlo methods and especially suited
for the generalization to larger systems, since it is easily implementable.
Fig.~\ref{fig-breakdown} shows the quasi-classical expectation value of the
Bloch vector (dashed green line) calculated by the simple phase space average
(\ref{eqn-pq-exvalues}) in comparison to the quantum Bloch vector
$\langle \hat{\vec J}\rangle/N$. One observes an excellent agreement. 
Note that the expectation values calculated from the classical distribution 
function and the phase space ensemble are indistinguishable on this scale
and thus only one curve is shown in the figure.

\subsection{Description of chaos in a driven two-mode system}

\begin{figure}[t]
\centering
\includegraphics[width=4cm,  angle=0]{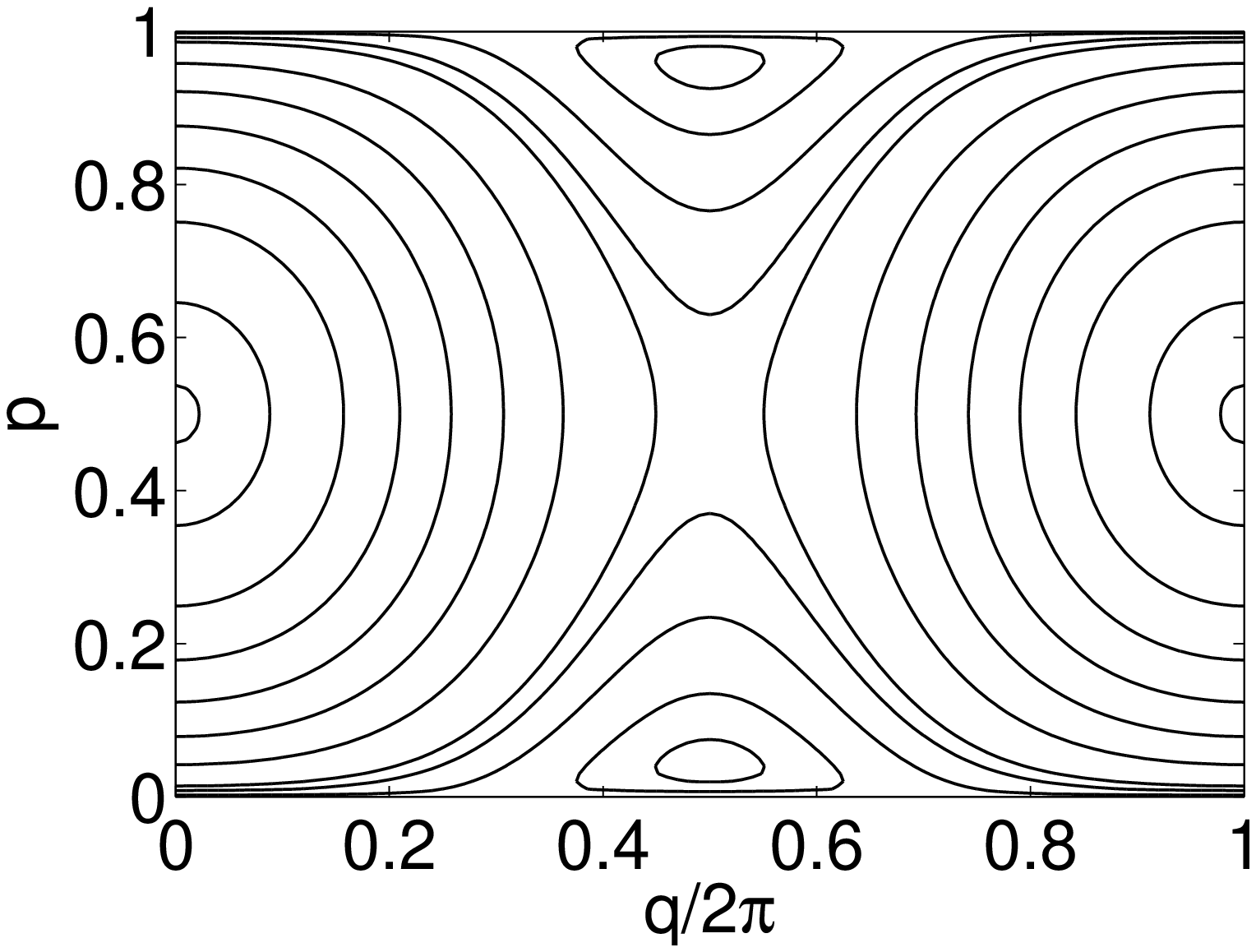}
\includegraphics[width=4cm,  angle=0]{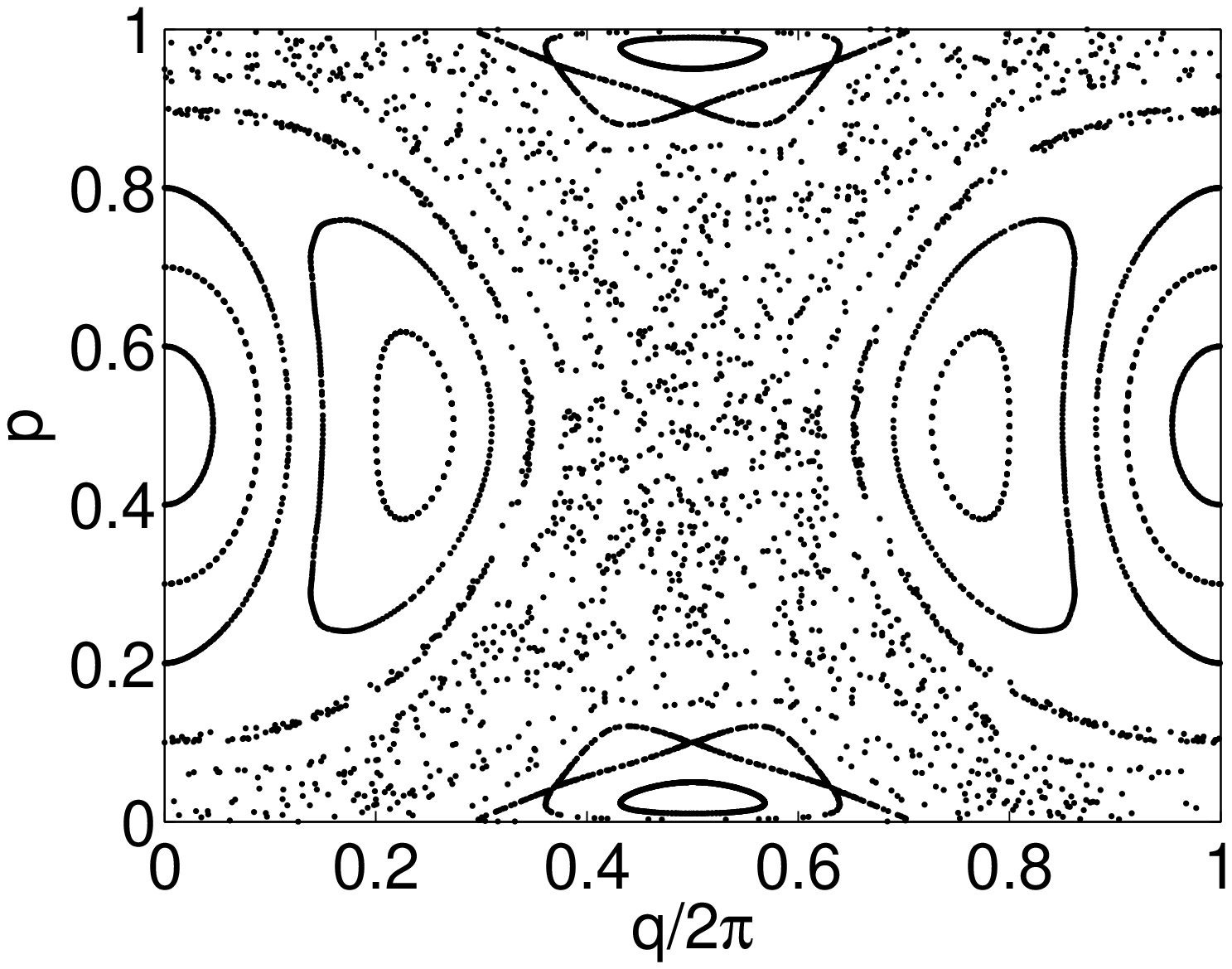}
\includegraphics[width=4cm,  angle=0]{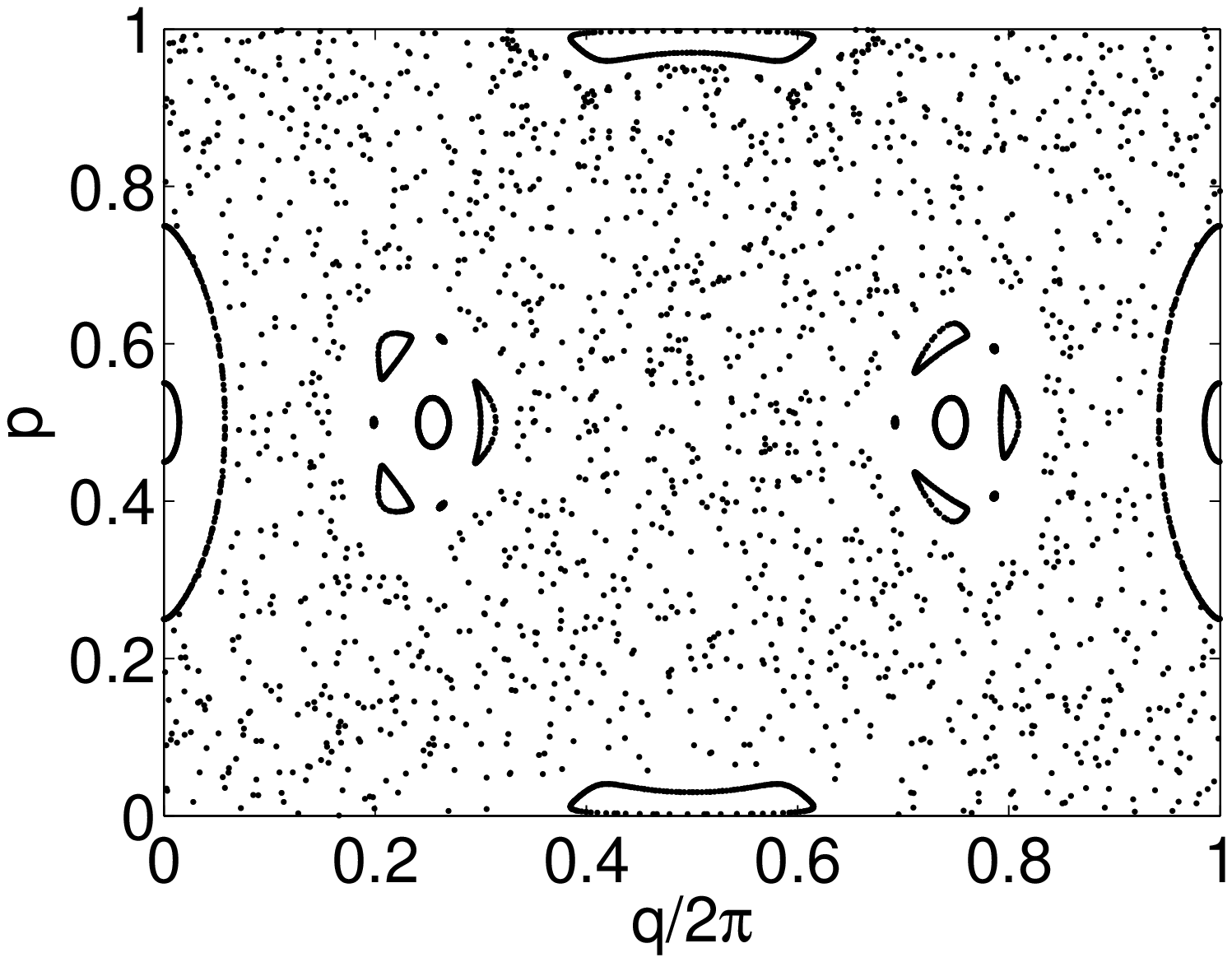}
\caption{\label{fig-driven-strobo}
Stroboscopic Poincar\'e section of the classical dynamics of a driven
two-mode BEC for $\Delta_0 = 1$, $g=5$, $\omega = 2\pi$ and 
$\Delta_1 = 0,0.2,0.5$ (from top left to bottom).}
\end{figure}
\begin{figure*}[t]
\centering
\includegraphics[width=16cm,angle=0]{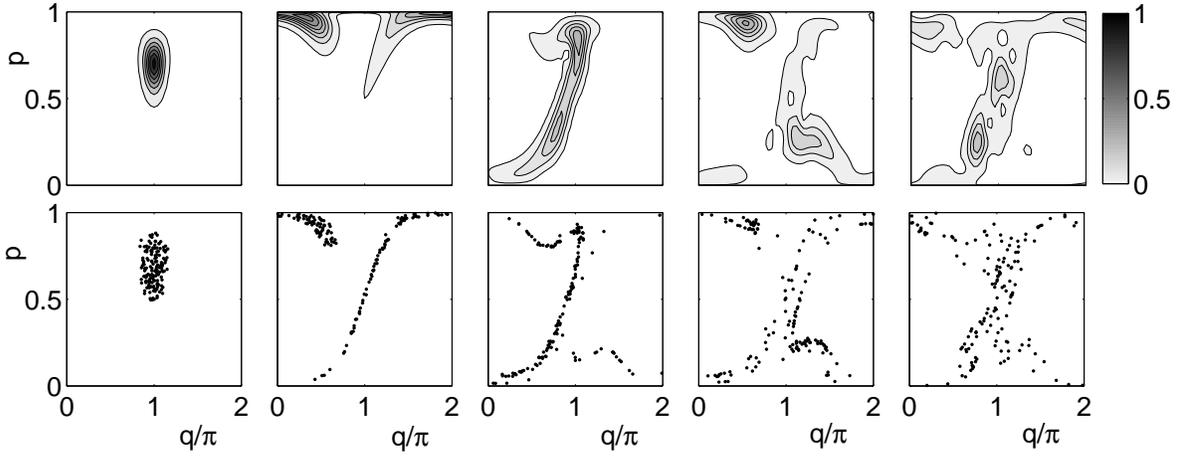}
\caption{\label{fig-driven-compare}
Comparison of quantum and classical dynamics in phase
space in the chaotic sea for $UN = 5$, $\Delta_0=1$, $\omega = 2\pi$ and 
$\Delta_1 = 0.5$ and $N=20$ particles. Upper panels: Exact Evolution of 
the Husimi density at times $t=0,1,2,3,4$ (from left to right).
Lower panels: Dynamics of an ensemble of 150 classical trajectories.}
\end{figure*}

\begin{figure}[t]
\centering
\includegraphics[width=8cm,  angle=0]{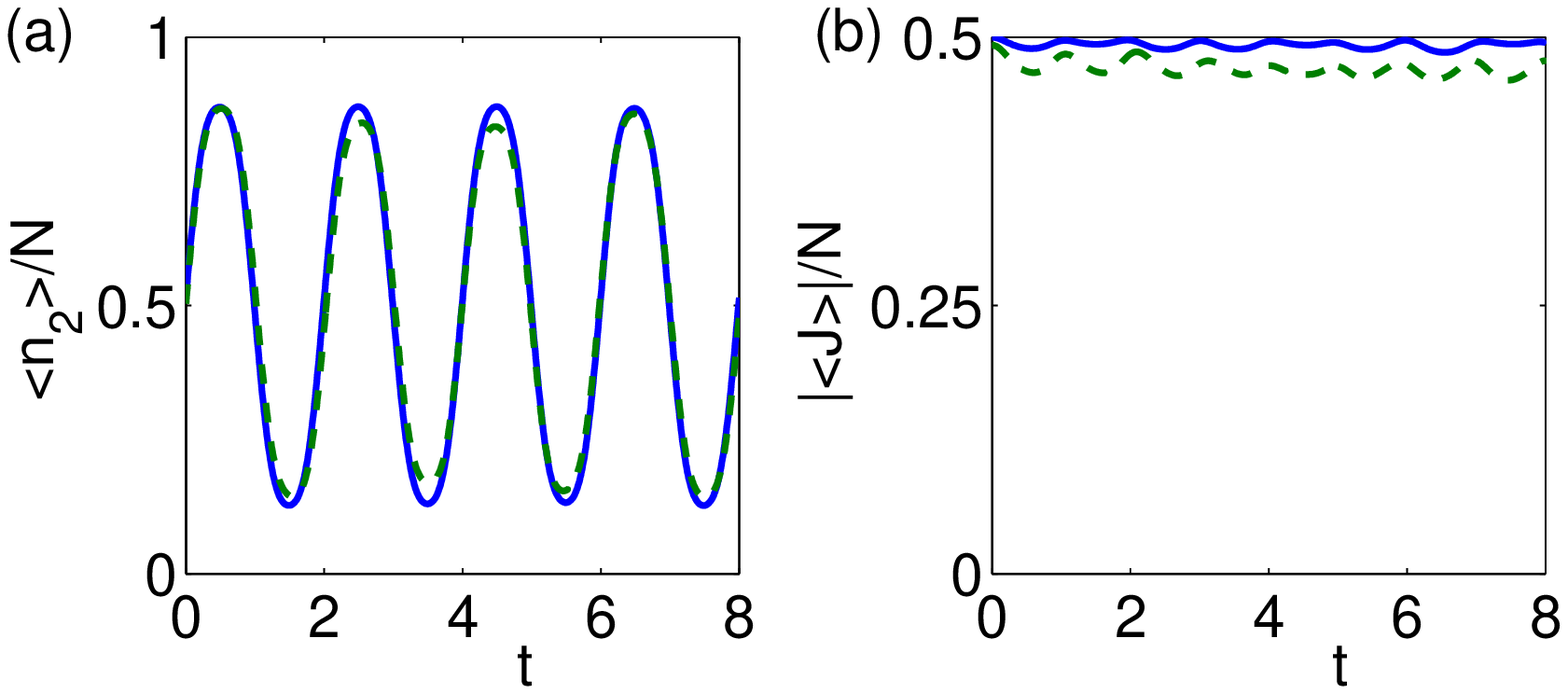}
\includegraphics[width=8cm,  angle=0]{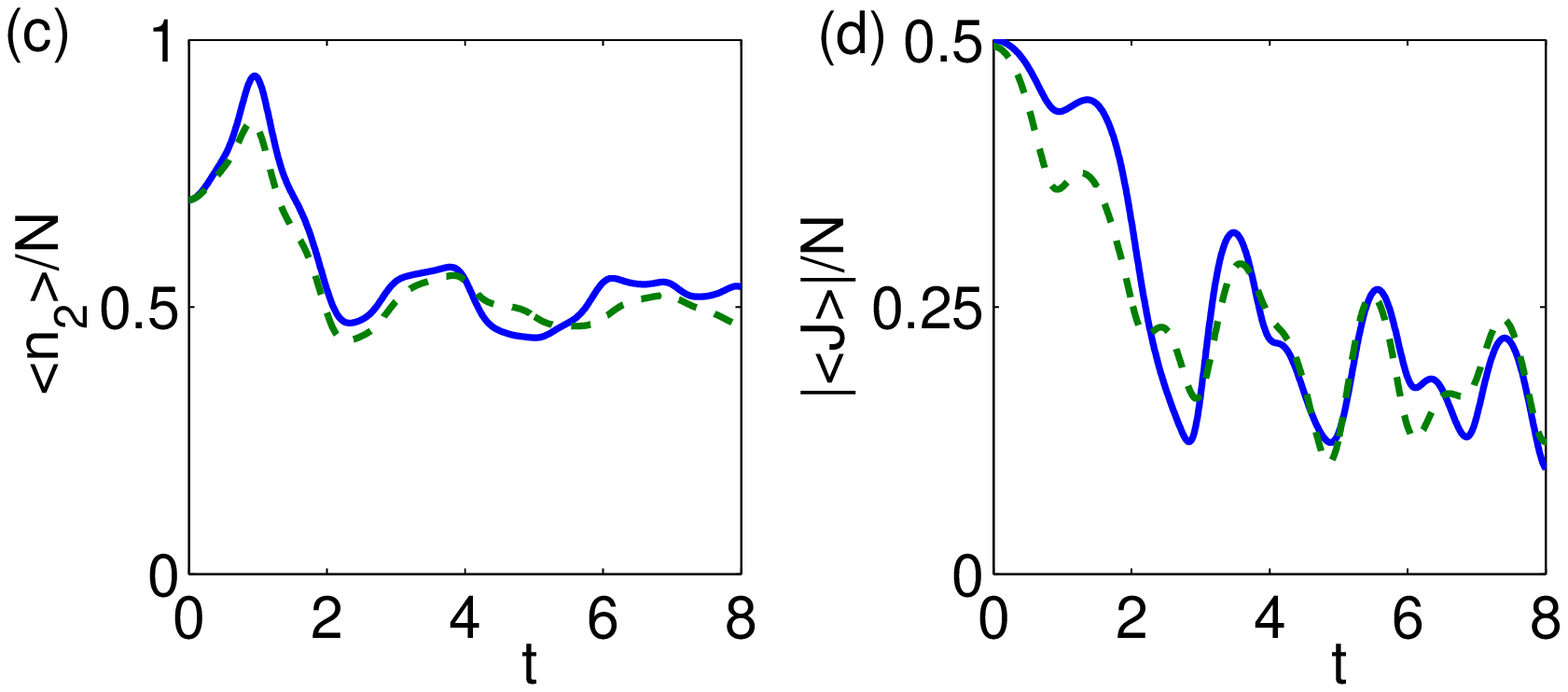}
\caption{\label{fig-driven-dyn}
(Color online) Quantum dynamics in a driven double-well trap for an initially coherent 
state localized at $(p,q) = (1/2,3\pi/2)$ on a regular island (a),(b)
and localized at $(p,q) = (0.7,0)$ in the chaotic sea (c),(d).
The population in the second well $\langle \hat n_2(t) \rangle/N$  
is plotted on the left-hand side and the magnitude of the Bloch vector
$|\langle \hat{ \vec J(t)} \rangle|/N$ is plotted on the right-hand 
side, both calculated exactly (solid blue lines) and with a classical
ensemble simulation (dashed green lines). 
The remaining parameters are $UN = 5$, $\Delta_0=1$, $\omega = 2\pi$,
$\Delta_1 = 0.5$ and $N=50$ particles.
}
\end{figure}

Periodically driven Bose-Hubbard systems are of current interest 
(see, e.g., \cite{Zhon08} and references therein).
In this paragraph, we consider the dynamics in a double-well trap for 
a time-dependent tunneling matrix element
\be
  \Delta(t) = \Delta_0 + \Delta_1 \cos(\omega t).
\ee
Such a driving can be implemented easily in an optical setup by 
varying the intensity of the laser beams forming the optical lattice.
Figure~\ref{fig-driven-strobo} shows a stroboscopic Poincare section of 
the mean-field dynamics for the parameters $\Delta_0 = 1$, $g=5$, 
 $\Delta_1 = 0, 0.2, 0.5$ and $\omega = 2\pi$ so that the period
is $T=1$. The fixed points of 
the stroboscopic mapping correspond to the nonlinear Floquet solutions 
of the driven GPE (cf. also \cite{Zhon08}).
For $\Delta_1 = 0$ one obviously recovers the undriven system, which is 
completely regular. A mixed regular-chaotic dynamics is found when $\Delta_1$ 
is increased, the region around the hyperbolic fixed point $(p,q) = (1/2,\pi)$ 
being the first to become chaotic.
Fig.~\ref{fig-driven-strobo} (b) shows the phase space for $\Delta_1 = 0.5$,
where only tiny regular islands have survived, among them regions around 
the symmetric fixed point $(p,q) = (1/2,0)$ and the self-trapping fixed points.
The regular islands around $(p,q) = (1/2,\pi/2)$ and $(p,q) = (1/2,3\pi/2)$
correspond to a period doubled fixed point. They are of particular interest
since they support regular oscillations of the population with an extremely
large amplitude as illustrated in Fig.~\ref{fig-driven-dyn}.

If the classical dynamics is unstable, an initially coherent quantum 
state will be strongly distorted. For an isolated unstable fixed point, 
the state spreads along the unstable manifold while it is squeezed 
in the orthogonal direction as shown in Fig.~\ref{fig-zusammenbruchps}.
This makes the standard mean-field approximation fail, but may even be 
desirable \cite{Sore01}. Classical chaos has a much more dramatic effect: 
An initial coherent state will rapidly spread over 
the entire chaotic sea. Thus, the description by a sharply peaked distribution
instantaneously fails and the Bogoliubov approach breaks down immediately.
Again this breakdown is resolved by the introduction of a phase space
ensemble as shown in Fig.~\ref{fig-driven-compare}. The classical 
ensemble well describes the spreading of the wave packet in the chaotic 
sea, demonstrating the power of the phase space approach also for 
classically chaotic dynamics.

The differences between regular and chaotic classical dynamics is furthermore
illustrated in Fig.~\ref{fig-driven-dyn}, where we have plotted the dynamics 
of the population in the second well $\langle \hat n_2(t) \rangle/N$ and 
the magnitude of the Bloch vector $|\langle \hat{ \vec J}(t) \rangle|/N$ for 
two different initial states. The figures compare the exact many-particle
dynamics (solid blue line) to the results of an ensemble simulation 
(dashed green line), showing a very good agreement for both regular and 
chaotic dynamics. If the initial state is located on an island 
as in Fig.~\ref{fig-driven-dyn} (a) and (b), the population imbalance 
oscillates regularly and the magnitude of the Bloch vector remains close 
to its maximum value $|\langle \hat{\vec J} \rangle| = N/2$. In this example 
we have chosen an initial state localized on a period-double fixed point 
(cf.~Fig.~\ref{fig-driven-strobo}), so that the population difference 
oscillates with only half of the driving period. The dynamics of a state 
initially localized in the chaotic sea is shown Fig.~\ref{fig-driven-dyn} 
(c) and (d). Oscillations are rapidly damped out and the magnitude of the 
Bloch vector drops to very small values.

\begin{figure}[t]
\centering
\includegraphics[width=8cm,  angle=0]{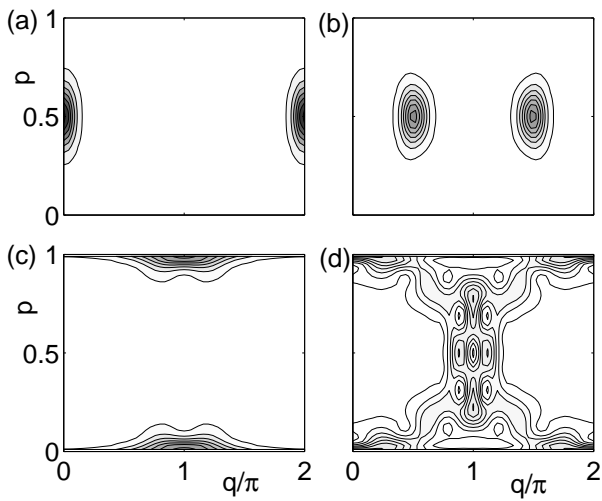}
\caption{\label{fig-driven-floquet}
Husimi representation of four typical Floquet states for a driven
double-well trap with parameters $UN = 5$, $\Delta_0=1$, $\omega = 2\pi$,
$\Delta_1 = 0.5$ and $N=50$ particles.
}
\end{figure}

As mentioned above, the classical phase space structure does not only 
affect the quantum dynamics, but also the organization the eigenstates
(cf.~Sec.~\ref{sec-phasespace}).
Due to the mixed phase space, this leads to much richer structures in the 
driven case. Instead of the eigenstates of the Hamiltonian, we now have
to consider the Floquet states, which are defined as the eigenstates of 
the time evolution operator over one period $T$,
\be
  \hat U(T) = \hat{\cal T} \exp \left( -i \int_0^T \hat H(t') dt' \right),
\ee
where $\hat{\cal T}$ is the time ordering operator. Figure~\ref{fig-driven-floquet} 
shows the Husimi representation of four typical Floquet states for $\Delta_1 = 0.5$. 
Three regular states are shown in the parts (a)--(c), corresponding to the 
symmetric, the self-trapping and the period-doubled fixed point, respectively.
The period-doubled Floquet state shown in part (c) oscillates with 
a large amplitude during one driving period, in which it remains well 
localized. In contrast, the chaotic eigenstate are delocalized over the 
complete chaotic sea -- a typical example is shown in part (d) of the figure.

\section{Beyond mean-field: Depletion and heating of the condensate}
\label{sec-beyond}

In this section we show that the Liouville picture also 
provides an excellent quasi-classical approximation of many-particle quantities
that are not accessible within the common single-trajectory 
mean-field description, since it also takes into account approximately 
the higher moments of the quantum state. 
In particular one can readily determine the deviation from a pure BEC 
and calculate the higher moments of the angular momentum operators
(\ref{eqn-angular-op}) that can be used to quantify many-particle 
entanglement \cite{Sore01}. In this paragraph we will especially discuss
the first issue for the two- and three-mode case in more detail. This is 
closely related to the preceding sections, as it is a clear signature
of the regular and chaotic dynamics in the corresponding quantum system. 

However, the Liouville description is not restricted to standard Bose-Hubbard 
systems. Therefore, we consider the generalization of the two-mode system to an open 
system in the last paragraph of this section. Here, one quantity of particular 
interest is the coherence factor between the two modes which decreases during
the heating process \cite{Gati06}. It is shown that the Liouville description 
provides a valuable approximation of this decoherence process.

\subsection{The two-mode Bose-Hubbard system}

\begin{figure}[t]
\centering
\includegraphics[width=8cm, angle=0]{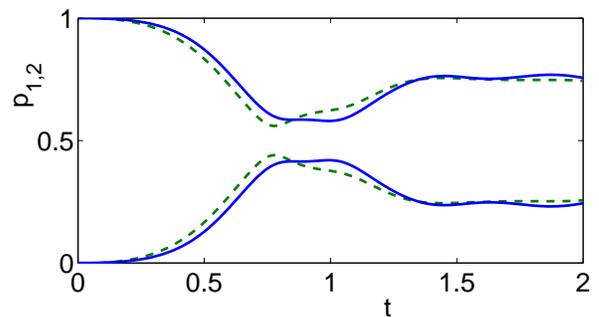}
\caption{\label{fig-breakdown_ev}
(Color online)  Eigenvalues of the reduced single-particle density matrix 
(\ref{eqn-spdm}) calculated from the quantum (solid blue line) and the 
quasi-classical (dashed green line) Bloch vector depicted in 
Fig.~\ref{fig-breakdown}.}
\end{figure}

The deviation from a pure BEC can be characterized by the eigenvalues of the reduced
single-particle density matrix (SPDM)
\be
  \rho_{\rm red} = \frac{1}{N} \left( \begin{array}{c c}
   \langle \hat a_1^\dagger \hat a_1 \rangle & \langle \hat a_1^\dagger \hat a_2 \rangle \\
   \langle \hat a_2^\dagger \hat a_1 \rangle & \langle \hat a_2^\dagger \hat a_2 \rangle \\
   \end{array} \right),  \label{eqn-spdm}
\ee
which are easily found to be given by
\be
  \lambda_{\pm} = \frac{1}{2} \pm \frac{|\langle \hat{\vec J} \rangle|}{N}. 
\ee
The leading eigenvalue of the SPDM gives the condensate fraction, i.e. the 
relative population of the condensate mode \cite{Legg01}. Thus, a BEC is 
pure if this eigenvalue is equal to one, i.e.~if the Bloch vector lies on the surface 
of the Bloch sphere, $|\langle \hat{\vec J} \rangle|/N = 1/2$. 
As discussed above, the expectation value of the angular momentum operators
can be calculated from the quasi-classical phase space distribution function 
$\rho(p,q)$ in a good approximation. From these it is also possible to 
reconstruct the SPDM and its eigenvalues completely classically. Furthermore, 
it is important to note that these quantities are not accessible within a 
single trajectory mean-field approach, which always assumes a pure BEC and thus guaranties
$|\langle \vec s \rangle|= 1/2=const$. 
To illustrate our point further we compare the eigenvalues of the SPDM from the 
quasi-classical Liouville approximation with the exact quantum results in 
Fig.~\ref{fig-breakdown_ev}.

Initially one eigenvalue is unity, indicating a pure BEC equivalent to a 
product state with every particle in the same mode. This eigenvalue decreases 
rapidly when the Husimi function approaches the hyperbolic fixed 
point (cf.~Fig.~\ref{fig-zusammenbruchps}), indicating a rapid depletion of 
the condensate mode. One observes that the classical calculation reproduces 
this depletion of the condensate mode very well. 

\subsection{The three-mode Bose-Hubbard system}

\begin{figure}[bt]
\centering
\includegraphics[width=7cm,  angle=0]{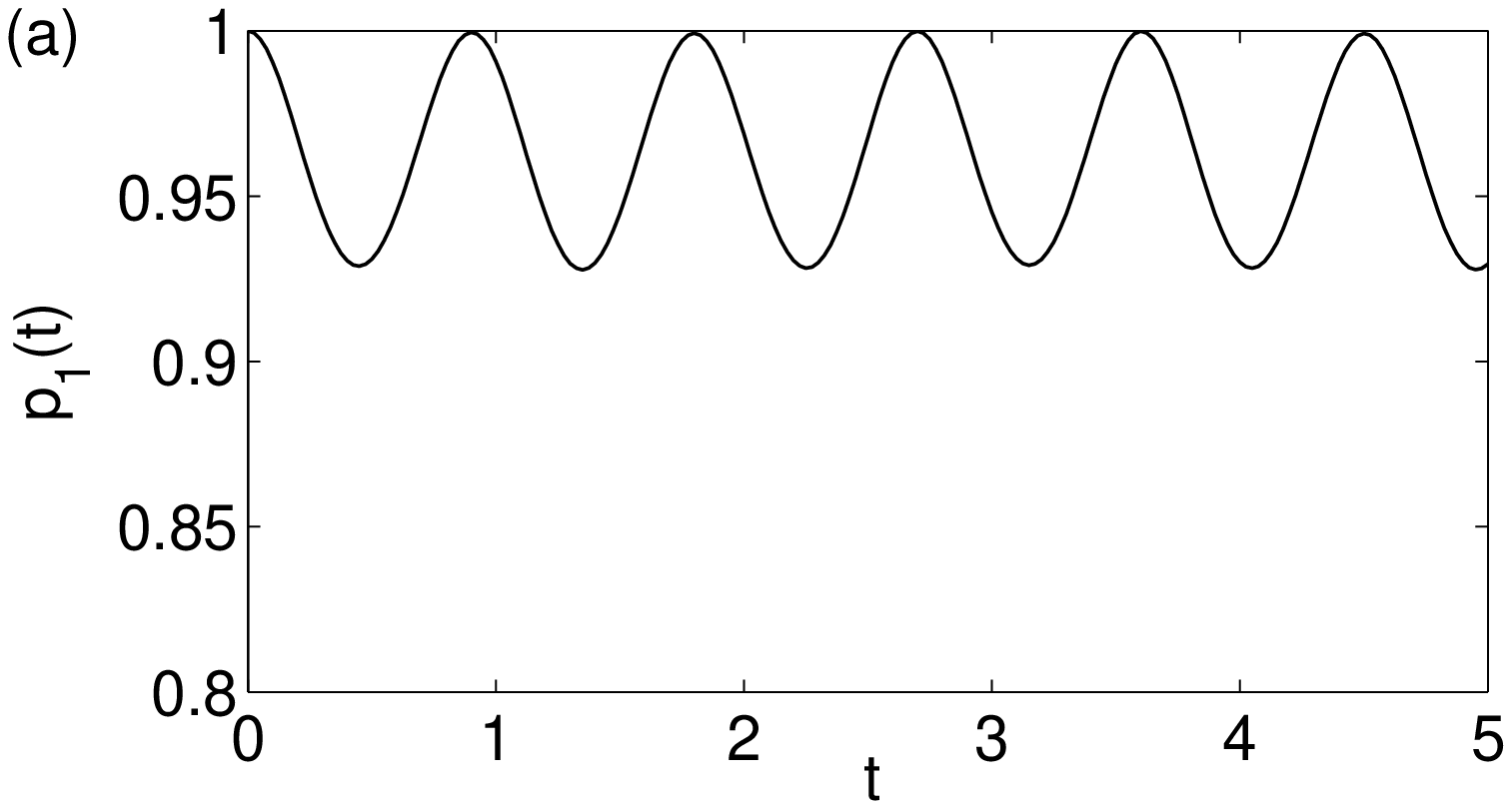}
\includegraphics[width=7cm,  angle=0]{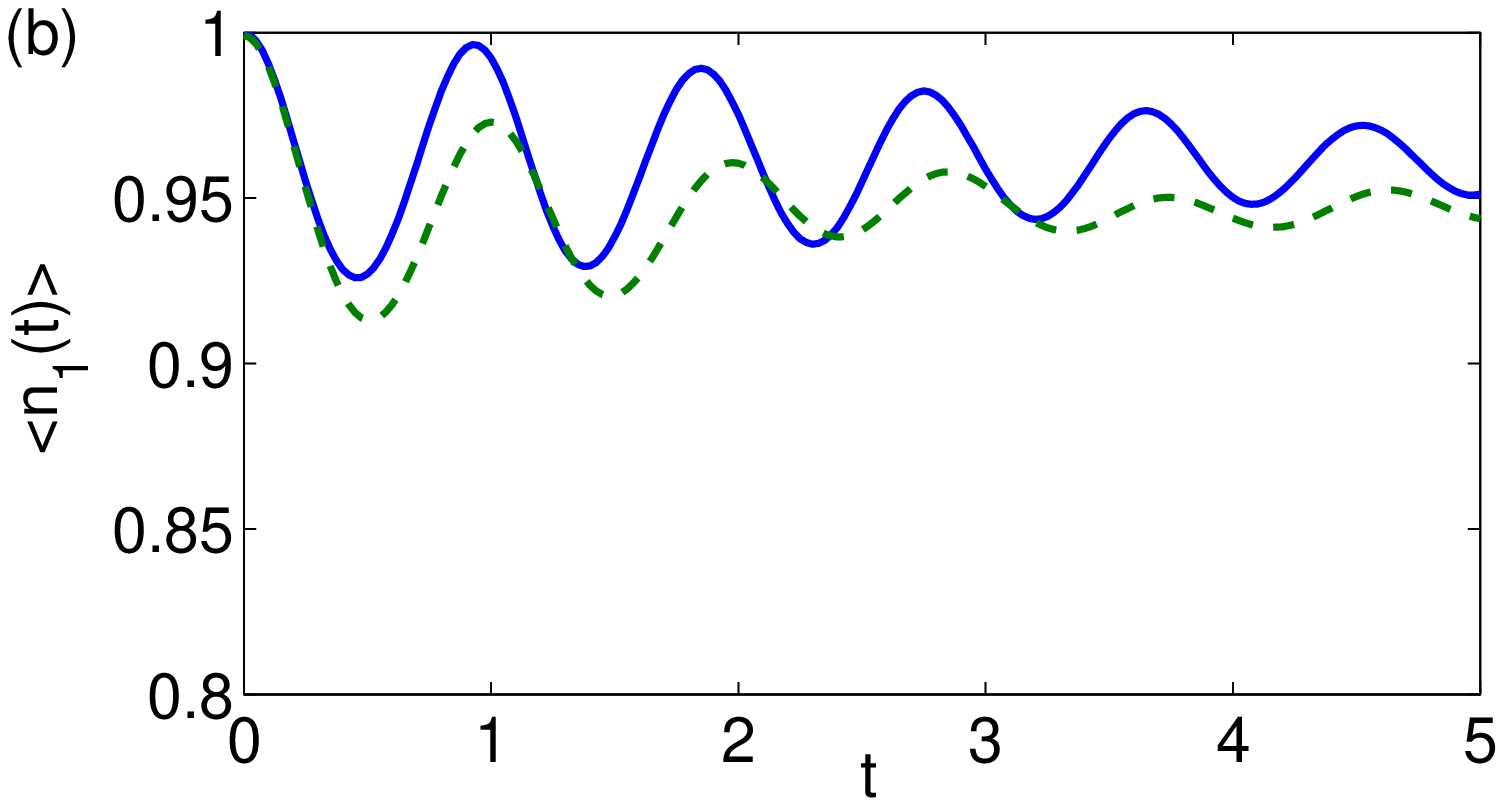}
\caption{\label{fig-3level_reg}
(Color online) Classical and quantum dynamics in a three-mode system for an initial 
state in the regular region of classical phase space ($\psi_1(0) = 1$).
The upper panel shows the occupation in the first well $p_1(t)$ calculated 
within the standard mean-field approximation. The lower panel shows the 
quantum expectation value $\langle \hat n_1(t) \rangle/N$ (solid blue line) 
and the ensemble expectation value (dashed green line). The parameters 
are $\epsilon_1 = 2$, $\epsilon_3 = 4$, $\Delta_{12} = \Delta_{23} = 1$, 
$UN=-10$ and $N=80$ particles.}
\end{figure}

\begin{figure}[t]
\centering
\includegraphics[width=7cm,  angle=0]{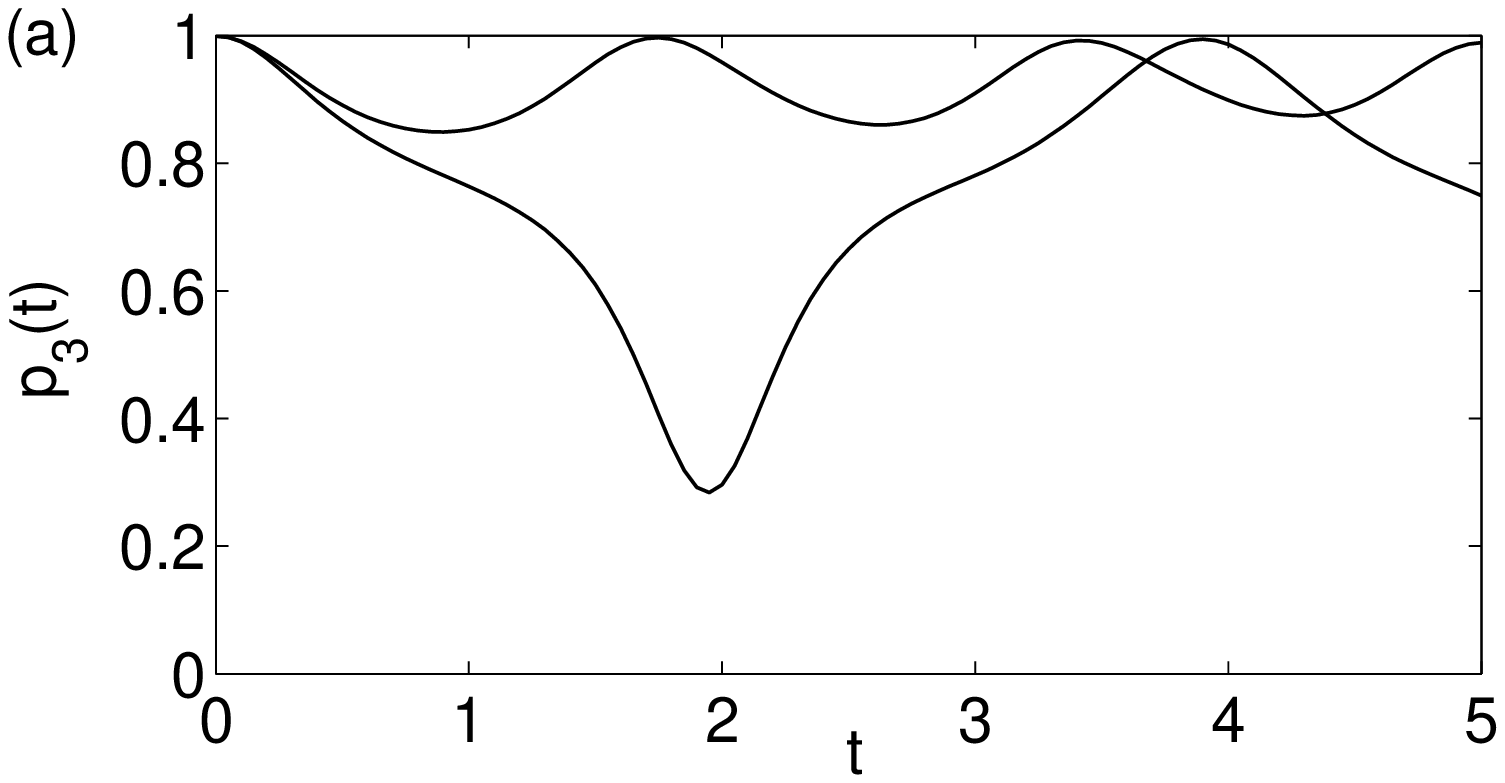}
\includegraphics[width=7cm,  angle=0]{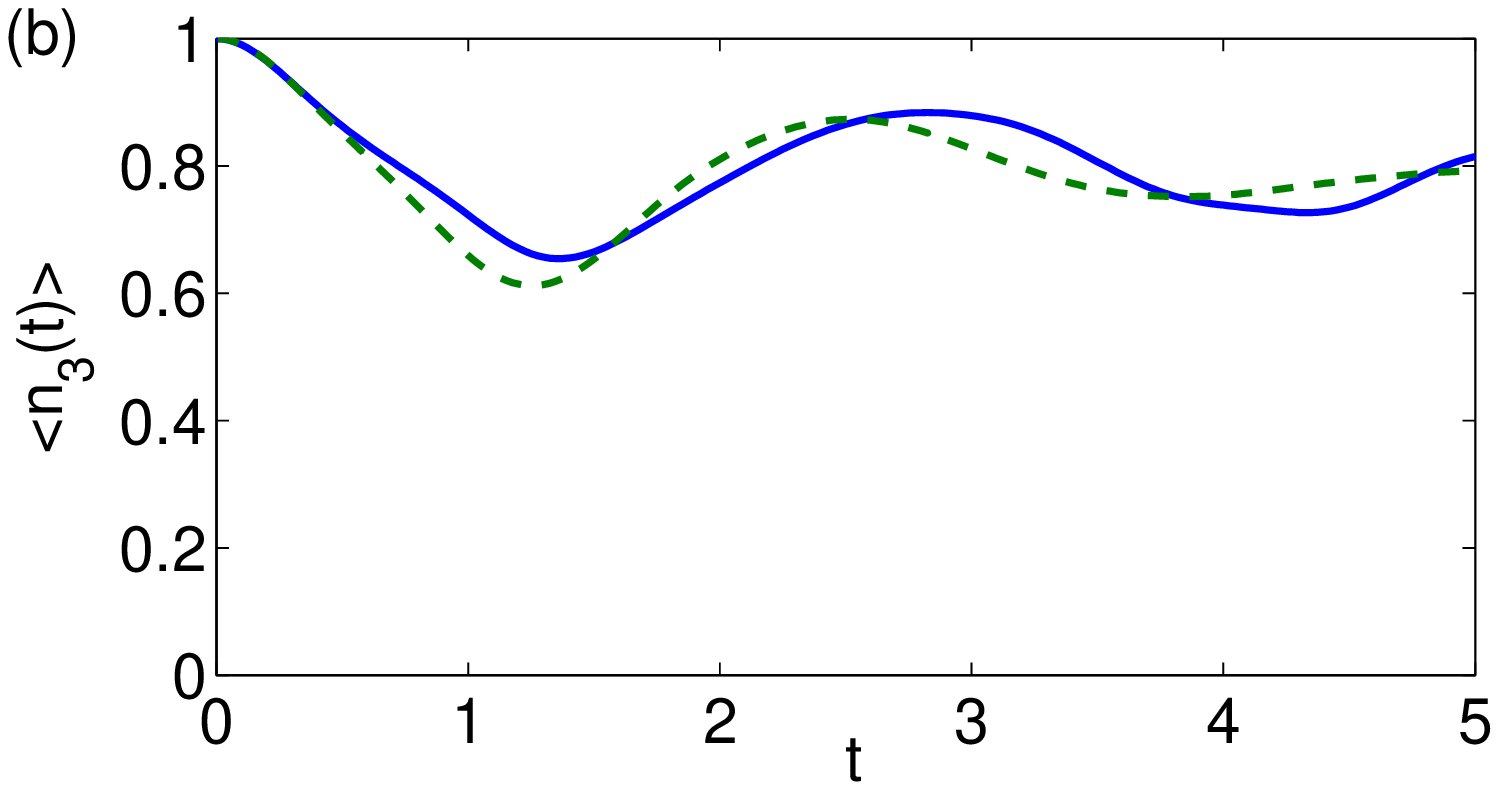}
\caption{\label{fig-3level_chaos}
(Color online)
Classical and quantum dynamics in a three-mode system for an 
initial state in the chaotic region of classical phase space. 
The upper panel shows the occupation in the third well $p_3(t)$ 
calculated within the standard mean-field approximation for 
$p_3(0) =1$ and $p_3(0) = 0.999$, respectively. The lower
panel shows the quantum expectation value $\langle \hat n_3(t) \rangle/N$ 
(solid blue line) and the ensemble expectation value (dashed green line) 
for $\psi_3(0) = 1$. The parameters are $\epsilon_1 = 2$, 
$\epsilon_3 = 4$, $\Delta_{12} = \Delta_{23} = 1$, $UN=-10$ and $N=80$ particles.}
\end{figure}

The same method can now be used to analyse the three-mode case.
Two examples are shown in Figs.~\ref{fig-3level_reg} and \ref{fig-3level_chaos},
where the standard mean-field approximation on the left-hand is compared 
to the results of a full quantum calculation and a classical ensemble
simulation. The figures show the population in the first and the third
mode, respectively.
Regular oscillations with a small amplitude are found for an initial state
localized in the first well ($\psi_1(0) = 1$, cf. Fig. \ref{fig-3level_reg}).
These oscillations are weakly damped in the quantum system due to a dephasing 
process. This damping is well reproduced by a classical ensemble calculations, 
while there is no damping at all in the standard mean-field approach (upper panel).

The dynamics is completely different for an initial state in the chaotic region 
of classical phase space as shown in Fig.~\ref{fig-3level_chaos}.
The upper panel shows the mean-field trajectories for two slightly different initial
states ($p_3(0) =1$ and $p_3(0) = 0.999$). They deviate fast and differ 
completely for $t \geapprox 0.5$. 
The dynamics of the  corresponding quantum system is shown in the lower panel
 for an initial state which is completely localized in the third mode,
\be
  \ket{\Psi(0)} = \frac{1}{\sqrt{N!}} (\hat a_3^\dagger)^N \ket{0,0,0} \, .
\ee
The oscillations of the density $\langle \hat n_3 \rangle$ are rapidly damped 
out. Again this is well reproduced by a classical ensemble calculation.
In this case the only signature of chaos in the common mean-field description 
is the sensible dependence on the initial state, however again there is no 
indication of the damping process.

\begin{figure}[t]
\centering
\includegraphics[width=7cm,  angle=0]{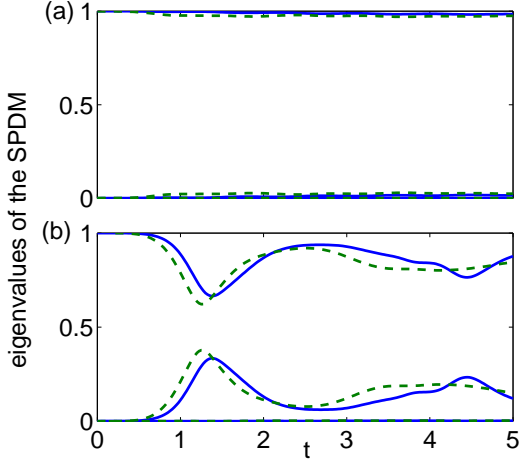}
\caption{\label{fig-3level_spdm}
(Color online) Evolution of the eigenvalues of the SPDM for (a) an initial state in the 
regular region of classical phase space ($\psi_1(0) = 1$) and (b) in 
the chaotic sea ($\psi_3(0) = 1$). Quantum results (blue solid 
lines) are compared to the classical ensemble estimate (dashed green
lines). The parameters are chosen as in Fig.~\ref{fig-3level_reg}
and \ref{fig-3level_chaos}, respectively.}
\end{figure}

As we have already seen in the preceding paragraph, the ensemble simulation 
gives a good estimate of many-particle quantities, which cannot be calculated 
in the standard mean-field theory. 
Figure~\ref{fig-3level_spdm} shows the evolution of the eigenvalues of the 
SPDM, which is defined analogously to Eqn.~(\ref{eqn-spdm}) for the 
two-mode case. The results from a numerical solution
of the many-particle dynamics are plotted as red lines, the ensemble
estimates as dashed blue lines. The leading eigenvalues 
of the SPDM gives the condensate fraction of the many-particle quantum state.
The condensate mode is depleted if the classical dynamics is unstable in
favor of the remaining modes \cite{Cast97,Cast98,Vard01b,Angl01}.
The differences between the regular and chaotic case are obvious. As expected, 
the condensate mode is significantly depleted if the dynamics is chaotic.
The leading eigenvalue decreases fast, whereas it remains close to one
if the dynamics is regular.
Furthermore one observes a very good agreement of the classical and the 
quantum results. Therefore the ensemble method provides a valuable tool 
to estimate the depletion of the condensate mode and to infer the nature of 
the many-body quantum state from a purely classical calculation. This
is especially useful for extended lattices, where quantum calculations 
become a hard problem. Another method to calculate the depletion of the 
condensate mode from the  mean-field dynamics was introduced in 
\cite{Cast97,Cast98}. However, its use is technically more involved 
than the simple ensemble approach, while it gives no additional information.

\subsection{Heating of a two-mode BEC}
\label{sec-heating}

Recently, the long-time dynamics of a BEC interacting with the background gas
has attracted a lot of interest \cite{Gati06,Gati06b,08mfdecay,08stores}. 
It was shown that collisions with 
the background gas lead to a decrease of the coherence of the two condensate 
modes, which can be used as a noise thermometer at extremely low temperatures.
In this section we want to discuss the heating of a BEC within the quasi-classical
phase space picture.

The main source of decoherence and heating is caused by elastic collisions with the 
background gas atoms and can described in leading order by the master equation
\cite{Angl97,Ruos98}
\be
  \dot{\hat \rho} &=& -\ri [\hat H,\hat \rho] - \frac{\gamma_1}{2} \sum_{j = 1,2}
  \hat n_j^2 \hat \rho + \hat \rho \hat n_j^2 - 2 \hat n_j \hat \rho \hat n_j.
   \label{eqn-master1}
\ee
The latter term describes elastic scattering events that conserve the number 
of particles in the condensate mode and only lead to phase decoherence (see, 
e.g., \cite{Wall85}). Thus they are readily described within the number 
conserving phase space approach discussed in the present paper, where the
interpretation as phase noise becomes especially clear.

The effects of the scattering events are conveniently understood and visualized
in quantum phase space. The evolution equation for the Husimi function is 
obtained as in Sec.~\ref{sec-phasespace}  by taking  the expectation value 
in $SU(2)$ coherent states:
\be
  \frac{\partial}{\partial t}  Q(p,q) &=&  \bra{p,q} \dot{\hat \rho} \ket{p,q}  \\
    &=& i \, \tr\left( \hat H \hat \rho - \hat \rho \hat H \ket{p,q}\bra{p,q} \right) \nn \\
    && \!\!\!\!\!\! - \frac{\gamma_1}{2} \sum_{j = 1,2} 
      \tr\left( \hat n_j^2 \hat \rho + \hat \rho \hat n_j^2 - 2 \hat n_j \hat \rho n_j \ket{p,q}\bra{p,q} \right). \nn 
\ee
Using the $\Dop^l$-algebra representation of the operators introduced above,
this equation can be cast into the form
\be
  \frac{\partial}{\partial t} Q(p,q) &=& -2 \, \Im(\Dop^\ell(\hat H)) Q(p,q) \\
      && \quad -  \frac{\gamma_1}{2} \sum_{j = 1,2} 
	\left( \Dop^\ell(\hat n_j) - \Dop^\ell(\hat n_j)^{*}  \right)^2
        Q(p,q) \nn \\
   &=& - \{\HH(p,q),Q(p,q)\} \nn \\
       && - 2 U p(1-p) \frac{\partial^2}{\partial p \partial q} Q(p,q) + \gamma_1 \frac{\partial^2}{\partial q^2} Q(p,q),\nn
    \label{eqn-heating-q}
\ee
with the classical Hamiltonian function (\ref{eqn-ham-clas}).
By an analogous calculation one finds the evolution equation for the Glauber-Sudarshan
distribution:
\be
  \frac{\partial}{\partial t} P(p,q) &=& - \{\HH(p,q),P(p,q)\} \\
      && + 2 U p(1-p) \frac{\partial^2}{\partial p \partial q} P(p,q) + \gamma_1 \frac{\partial^2}{\partial q^2} P(p,q), \nn
      \label{eqn-heating-p}
\ee
keeping in mind that the macroscopic interaction strength is now given by
$\tilde g = U(N+2)$ instead of $g = UN$.
In these representations the effect of the decoherence term $\sim \gamma_1$ becomes
most obvious: It leads to a diffusion of the relative phase of the two condensates
and thus to a blurring of the coherence of the two condensates modes. This effect
has been directly measured in the experiments of the Oberthaler group \cite{Gati06,Gati06b}.

\begin{figure}[tb]
\centering
\includegraphics[width=6cm, angle=0]{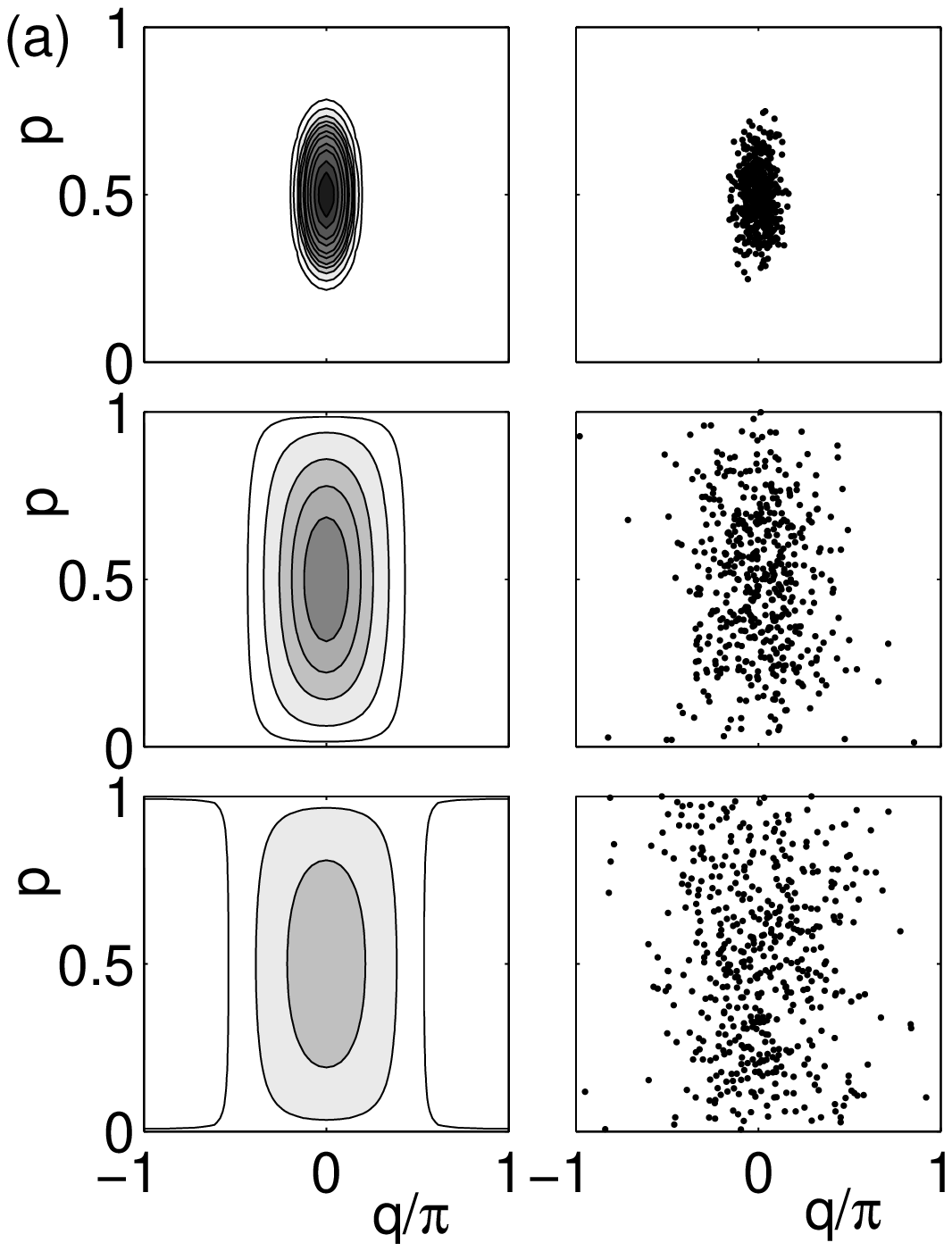}
\includegraphics[width=7cm, angle=0]{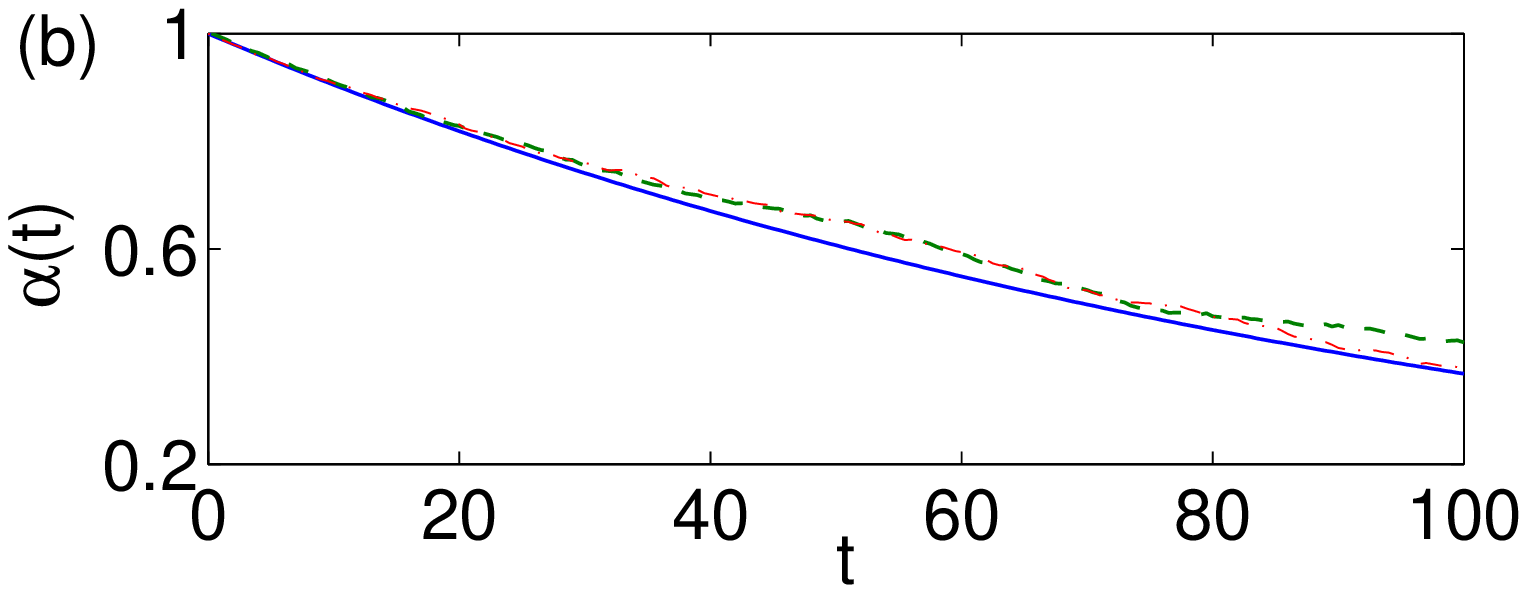}
\includegraphics[width=7cm, angle=0]{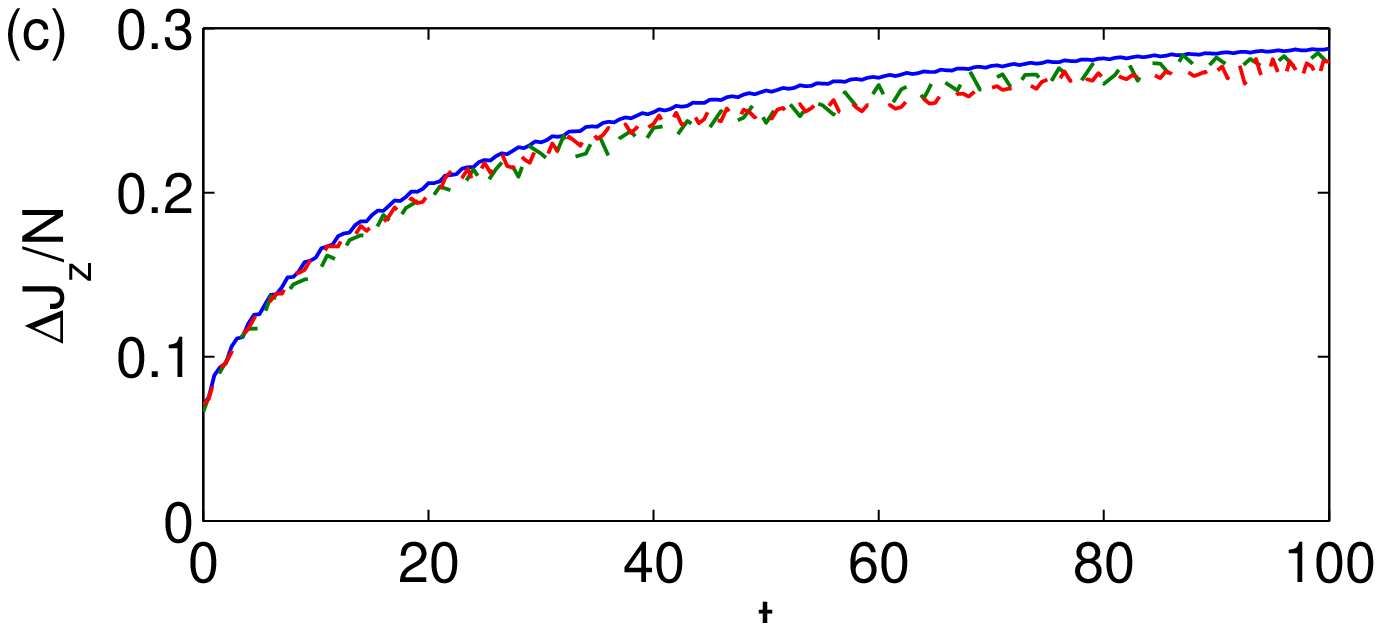}
\caption{\label{fig-heating-g0}
(Color online)
Heating of a two-mode BEC in quantum phase space for $UN = 0$ $\Delta = 1$, 
$\epsilon=0$ and $\gamma_1 = 0.01$. 
(a) Comparison of the exact quantum evolution of the Husimi function 
$Q(p,q,t)$ (left-hand side) and the stochastic dynamics of a classical 
phase space ensemble (right-hand side) for $t = 0,30,60$ (from bottom 
to top). Dark colors encode high values of the Husimi density.
(b) Decay of the coherence factor $\alpha(t)$. The exact 
quantum result (solid blue line) is compared to ensemble simulations 
based on the $Q$-function (dashed green line) and the $P$-function 
(dash-dotted red line).}
\end{figure}

If we neglect the quantum noise term $\sim g/N$, the equations 
(\ref{eqn-heating-q}) and (\ref{eqn-heating-p}) reduce to Fokker-Planck equations. 
Thus the condensate dynamics can again be interpreted in terms of phase space 
ensembles, now subject to the stochastic evolution equations
\be
  \dot p = - \frac{\partial \HH}{\partial q} \quad \mbox{and} \quad
  \dot q = + \frac{\partial \HH}{\partial p}  + \sqrt{2 \gamma_1} \xi(t),
  \label{eqn-heating-sde}
\ee
where $\xi(t)$ describes uncorrelated white noise:
\be
  \langle \xi(t) \rangle = 0  \quad \mbox{and} \quad 
  \langle \xi(t) \xi(t') \rangle = \delta(t-t').
\ee

\begin{figure}[tb]
\centering
\includegraphics[width=6cm, angle=0]{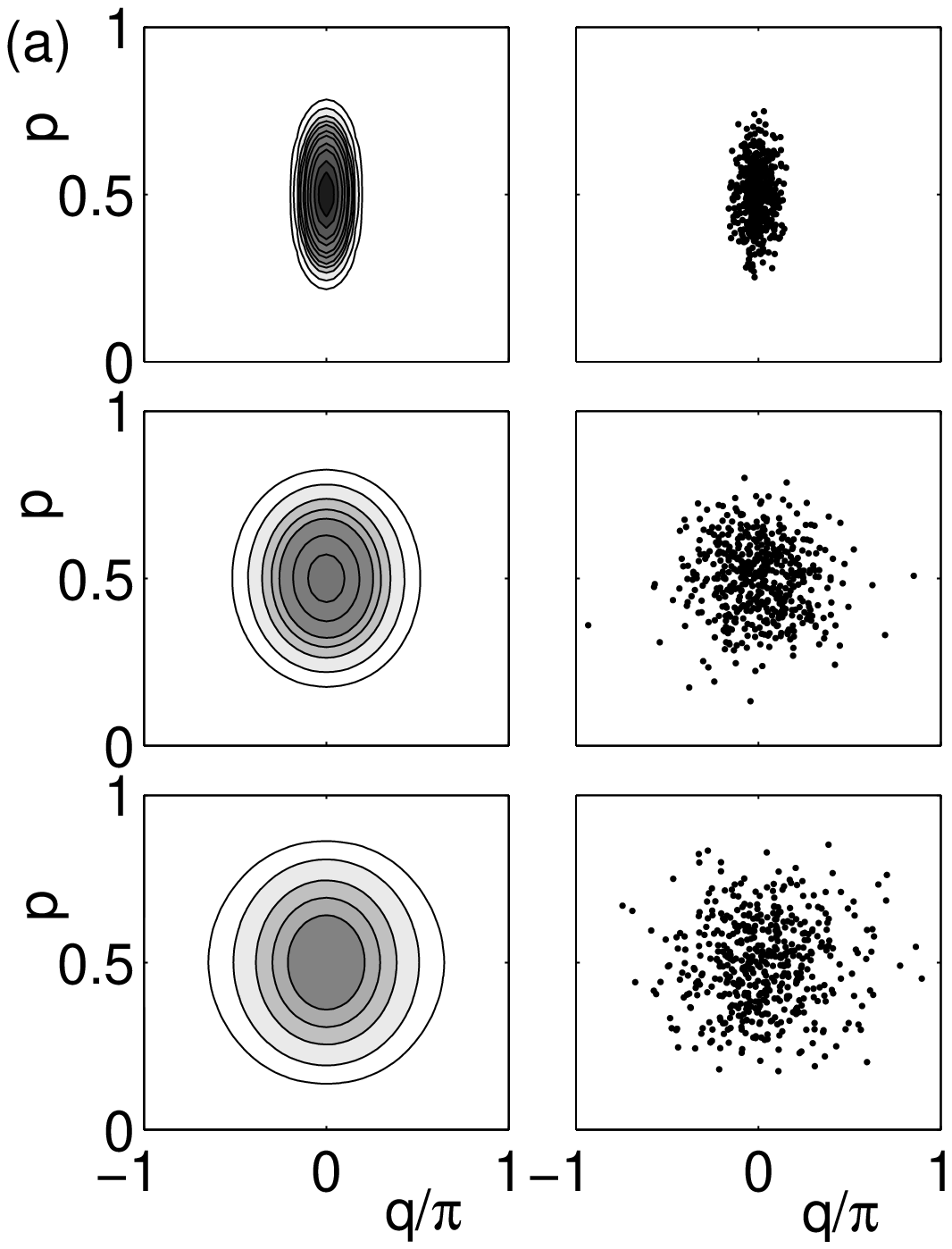}
\includegraphics[width=7cm, angle=0]{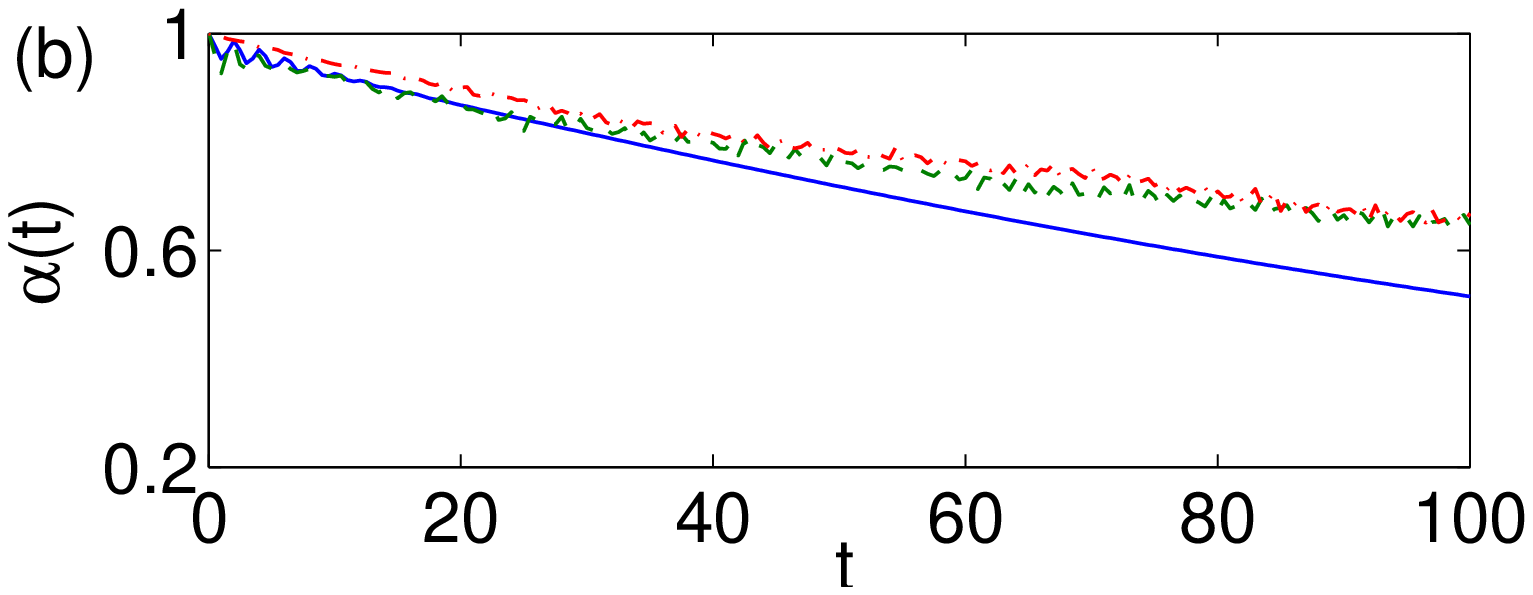}
\includegraphics[width=7cm, angle=0]{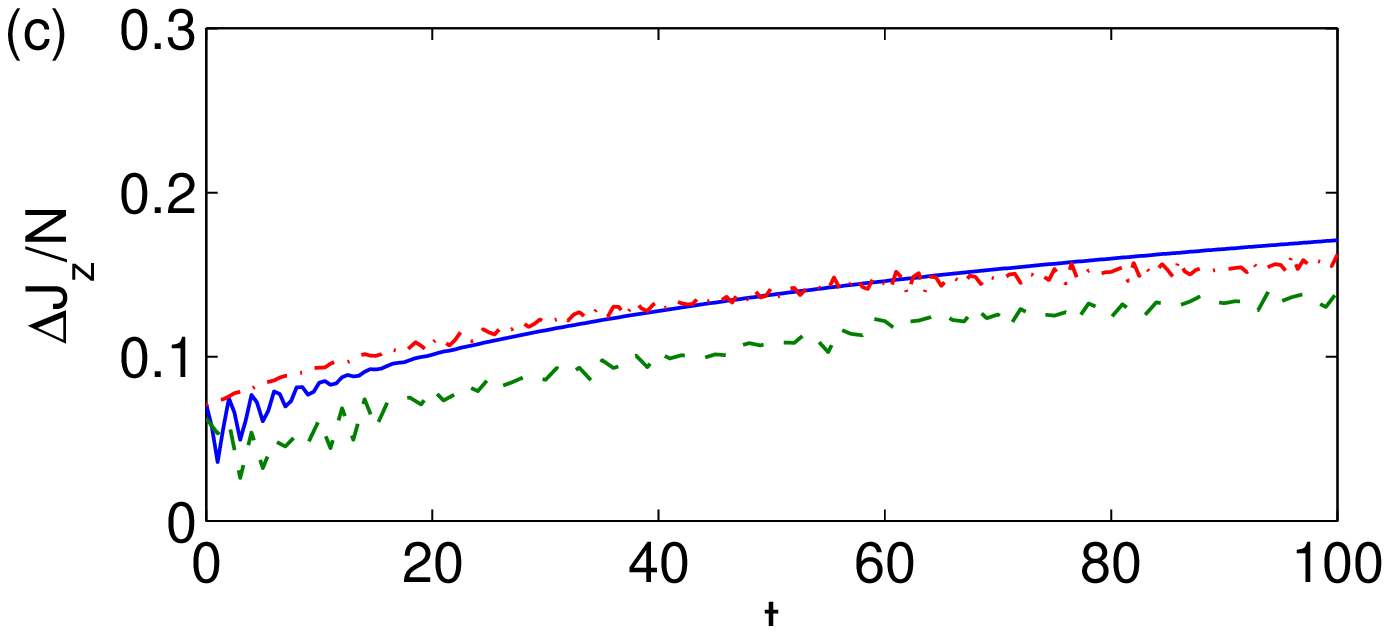}
\caption{\label{fig-heating-g10}
(Color online) As Fig.~\ref{fig-heating-g0}, however for $UN = 10$.}
\end{figure}

An example for this quasi-classical description of phase diffusion due to 
heating of the condensate is shown in Fig.~\ref{fig-heating-g0}(a) for 
$UN = 0$ and Fig.~\ref{fig-heating-g10}(a) for $UN = 10$, respectively. 
The left hand side shows the exact quantum evolution of the Husimi 
distribution $Q(p,q,t)$ calculated from the evolution of the density 
matrix $\hat \rho(t)$ according to the master equation 
(\ref{eqn-master1}). At $t=0$ the system is assumed to be in a coherent 
state $\ket{p =1/2, q = 0}$, i.e., a pure condensate with equal 
population and zero phase difference between the two wells. In the 
course of time, the Husimi function spreads and thus the coherence factor 
$\alpha(t) = \frac{2}{N} \langle \hat J_x \rangle_t$  decays as shown 
in part (b) of the figures.
The right-hand sides of Fig.~\ref{fig-heating-g0}(a) and
Fig.~\ref{fig-heating-g10}(a) show the evolution of a classical 
phase space ensemble initially distributed according to the Husimi 
function $Q(p,q,0)$. The single trajectories evolve according to 
the stochastic equations of motion (\ref{eqn-heating-sde}) and thus 
diffuse over the classical phase space. One observes that the quantum 
dynamics is well reproduced by the classical approach, especially the 
different shape of the Husimi function for $U = 0$ and $UN = 10$ after
the diffusion process.

For a more quantitative analysis we have plotted the coherence factor 
$\alpha(t)$ as well as the fluctuations of the population imbalance
in Fig.~\ref{fig-heating-g0} and Fig.~\ref{fig-heating-g10}.
The quantum result $\frac{2}{N} \langle \hat J_x \rangle_t$ is compared 
to the ensemble averages (\ref{eqn-pq-exvalues}) over $500$ classical 
trajectories for the $Q$-function and $P$-function, respectively.
In the linear case $UN = 0$ the mapping to stochastic evolution equations 
is exact and thus the deviations only result from the finite number of
classical representations. For $UN = 10$, however, quantum noise is 
neglected leading to a systematic underestimation of loss of phase 
coherence.

Furthermore one observes that the coherence factor $\alpha(t)$ decreases 
much slower for $UN = 10$ compared to the non-interacting case $UN = 0$
both in the exact and in the classical calculation. Similarly, the 
fluctuations of the population imbalance grow much slower in the interacting
case. 
These differences are readily understood from the structure of the classical 
phase space as shown in Fig.~\ref{fig-phasespace}.
One observes that the minimum of the classical Hamiltonian function 
$\HH(p,q)$ is much deeper for $UN = 10$, so that the classical 
trajectories are bound much stronger and phase diffusion is significantly 
reduced (cf. also \cite{GBJo07}). In addition, this leads to a stronger 
restriction of the phase space distribution in the $J_z$-axis.

\section{Comparison to other mean-field approaches}
\label{sec-comparison}

In the preceding sections we have introduced the number-conserving
phase space description of the Bose-Hubbard model. This approach 
provides a straightforward derivation of the (discrete) Gross-Pitaevskii 
(GP) equation in the macroscopic limit and allows to go beyond
mean-field theory. In the following we compare this method 
to other approaches to derive a mean-field approximation. 
The most common ones are the Bogoliubov and related approximations
which are discussed in detail in Sec.~\ref{sec-bogoliubov}. Phase space
methods are also frequently used, however starting from Glauber coherent 
states in almost all cases. A detailed comparison of the number conserving 
and the Glauber phase space description is given in 
Sec.~\ref{sec-comparison-phase}.
Furthermore  a considerable amount of work has been devoted to a rigorous derivation of
the GP energy functional and the GP dynamics in the macroscopic limit 
(see, e.g., \cite{Lieb00})

\subsection{Bogoliubov theory and related approaches}
\label{sec-bogoliubov}

The most common method to derive the (discrete) GP equation is the 
Bogoliubov approach, taking the expectation value of the Heisenberg 
equation of motion for the annihilation operator \cite{Bogo47}
\be
  i \frac{d}{dt} \langle \hat a_j \rangle = -J
  \left( \langle  \hat a_{j-1}  \rangle +  \langle  \hat a_{j+1}  \rangle  \right)
  + U \langle \hat a_j^\dagger \hat a_j \hat a_j \rangle
\ee 
and truncating the correlation functions
\be
  \langle \hat a_j^\dagger \hat a_k \hat a_\ell \rangle \approx
  \langle \hat a_j^\dagger \rangle \langle \hat a_k \rangle \langle \hat a_\ell \rangle.
  \label{eqn-bogo-trunc}
\ee
Quantum fluctuations are completely neglected in this equation.
They can be calculated approximately with the Bogoliubov-de Gennes
equations which linearize the equations of motion for $\hat a_j$
around the mean-field approximation given by the GP equation 
\cite{Gard97,Cast98}. 
If the initial state is a pure BEC, it is thus possible to calculate
expectation values to all orders and to infer the increase of the 
non-condensed atoms from the dynamical (in)-stability of the GP
equation \cite{Cast98}. However, this approximation includes no backreaction 
of the non-condensed atoms onto the condensate mode, so that this approximation 
is restricted to short times.

To overcome this problem, one has to truncate higher order expectations
values at a later stage than in Eqn.~(\ref{eqn-bogo-trunc}),
\begin{eqnarray}
  \langle \hat a_j^\dagger \hat a_j \hat a_j \rangle &\approx& 
  \langle \hat a_j^\dagger \rangle \langle \hat a_k \hat a_\ell \rangle
  + \langle \hat a_j^\dagger  \hat a_k \rangle \langle \hat a_\ell \rangle
  + \langle \hat a_j^\dagger  \hat a_\ell \rangle \langle \hat a_k \rangle \nn \\
  && \qquad - 2 \langle \hat a_j^\dagger \rangle \langle \hat a_k \rangle \langle \hat a_\ell \rangle,
\end{eqnarray}
and to derive equations of motion for the correlation functions
$\langle \hat a_j^\dagger \hat a_k  \rangle$ and 
$\langle \hat a_j \hat a_k \rangle$. Depending on the fact if 
and to which extend anomalous terms are neglected the resulting models are 
known under the name Hartree-Fock-Bogoliubov (HFB), HFB-Popov or 
Griffin (see, e.g., \cite{Grif96} and references therein).
However, all these methods face some characteristic difficulties
which are direct consequences of the breaking of the $U(1)$ 
symmetry \cite{Tikh07}. For instance, none of them conserves the 
total number of particles and, at the same time, allows for particle
exchange between the condensate and the remaining modes.

To overcome these problems, Anglin, Vardi and Tikhonenkov proposed
the Bogoliubov backreaction (BBR) method \cite{Vard01b,Angl01,Tikh07}.
In contrast to the $U(1)$ symmetry breaking approaches, only the number 
conserving operators $\hat E_{jk} = \hat a_j^\dagger \hat a_k$ are taken
into account. Again, equations of motion are calculated for the
expectation values $\langle \hat E_{jk} \rangle$ and the correlation 
functions $\langle \hat E_{jk}  \hat E_{\ell m} \rangle$, where higher
order expectation values are truncated. Several numerical examples
shown in \cite{Tikh07} suggest that the BBR method provides a better
approximation to the many-particle dynamics than HFB and is varieties. 

But still, BBR is limited to the first two moments of the operators 
$\hat E_{jk}$. Higher order correlation functions are not defined at 
all. This is clearly different within the phase space description,
which allows to approximate arbitrary momenta. 

From a practical viewpoint, it is thus fair to say that the phase space 
method embodies the best of both worlds: As the Bogoliubov-de Gennes 
approximation, it allows the calculation of expectation values to all 
orders and links the increase of the non-condensed atoms to the dynamical 
stability of the 'classical' GP equation. However, it also describes the 
depletion of the condensate mode and the long time dynamics just as the
HBF and the BBR method. Last but not least it is even simpler to use and 
implement then both of them. 

\subsection{Comparison to $U(1)$-symmetry breaking methods}
\label{sec-comparison-phase}

Quantum phase space descriptions are usually based on the celebrated
Glauber coherent states, which are superpositions of different number
states. In particular, the dynamics of ultracold atoms in optical lattices
has been discussed in \cite{Stee98}.
While this description is natural in quantum optics, where the birth 
and death of photons have to be taken into account, it does not take
into account the $SU(M)$ symmetry of the Bose-Hubbard Hamiltonian.
In this section, we will show that the number conserving description
is superior in describing a BEC with a fixed number of particles.
Note, however, that the $SU(M)$ symmetry is explicitly broken in the Gutzwiller 
approximation so that Glauber states provide the correct starting point in this case
\cite{Jain04}. In this approximation, one considers only a single 
lattice site which couples to the mean-field on the remaining lattice 
sites so that the local particle number is not conserved any longer.

The $U(1)$ symmetry breaking phase space description is based on the 
Glauber coherent states 
\be
  \ket{\alpha_1,\alpha_2} = e^{-(|\alpha_1|^2 + |\alpha_2|^2)/2} 
  \sum_{n_1,n_2 = 0}^\infty
  \frac{\alpha_1^{n_1} \alpha_2^{n_2}}{\sqrt{n_1! n_2!}} \ket{n_1,n_2},
  \label{eqn-glauberstate}
\ee
where the particle number follows a Poissonian distribution. Let us 
briefly discuss the relation between the phase space representations 
based on $SU(2)$ and Glauber coherent states.

\begin{figure}[tb]
\centering
\includegraphics[width=8cm, angle=0]{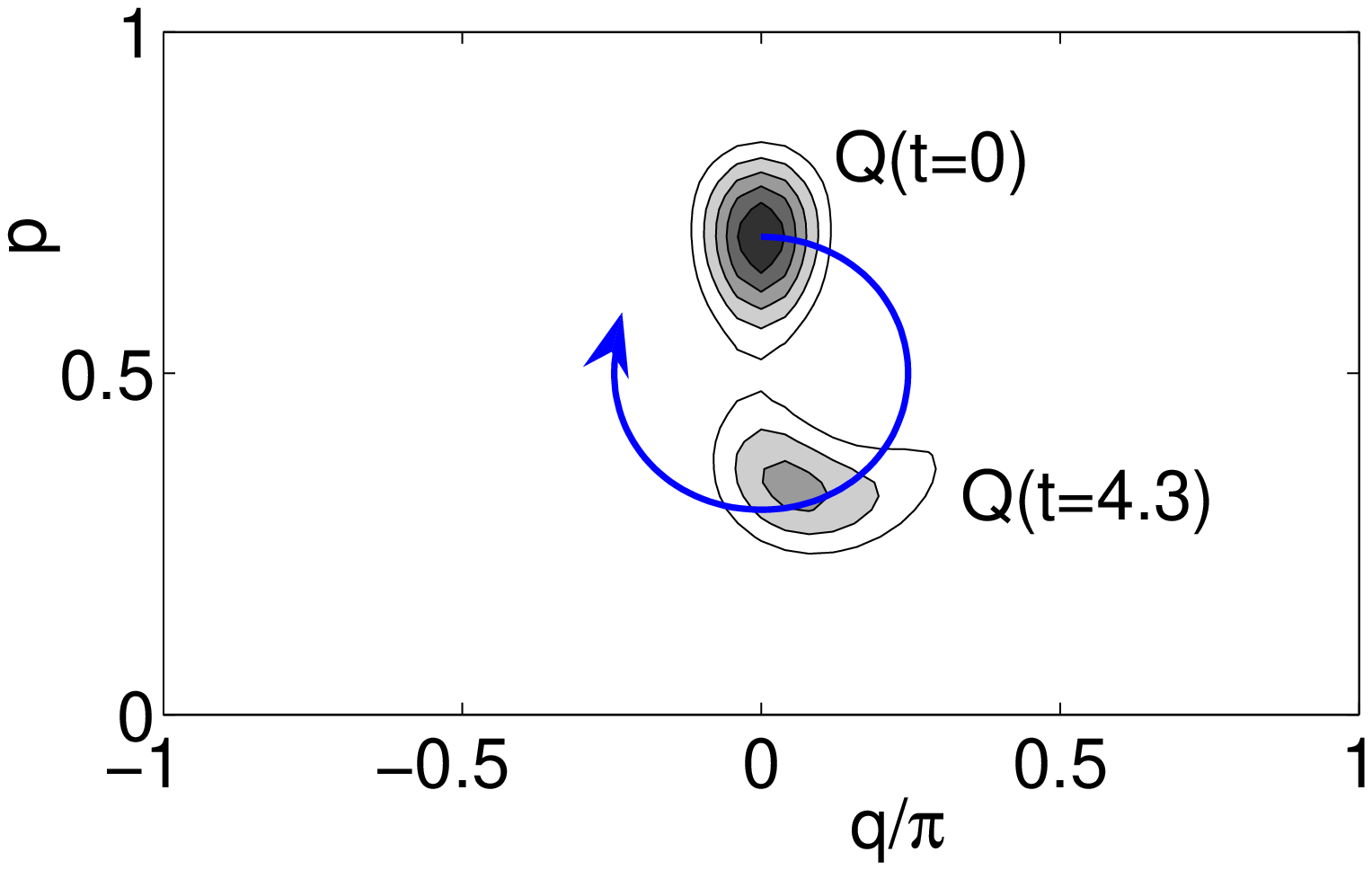}
\includegraphics[width=8cm, angle=0]{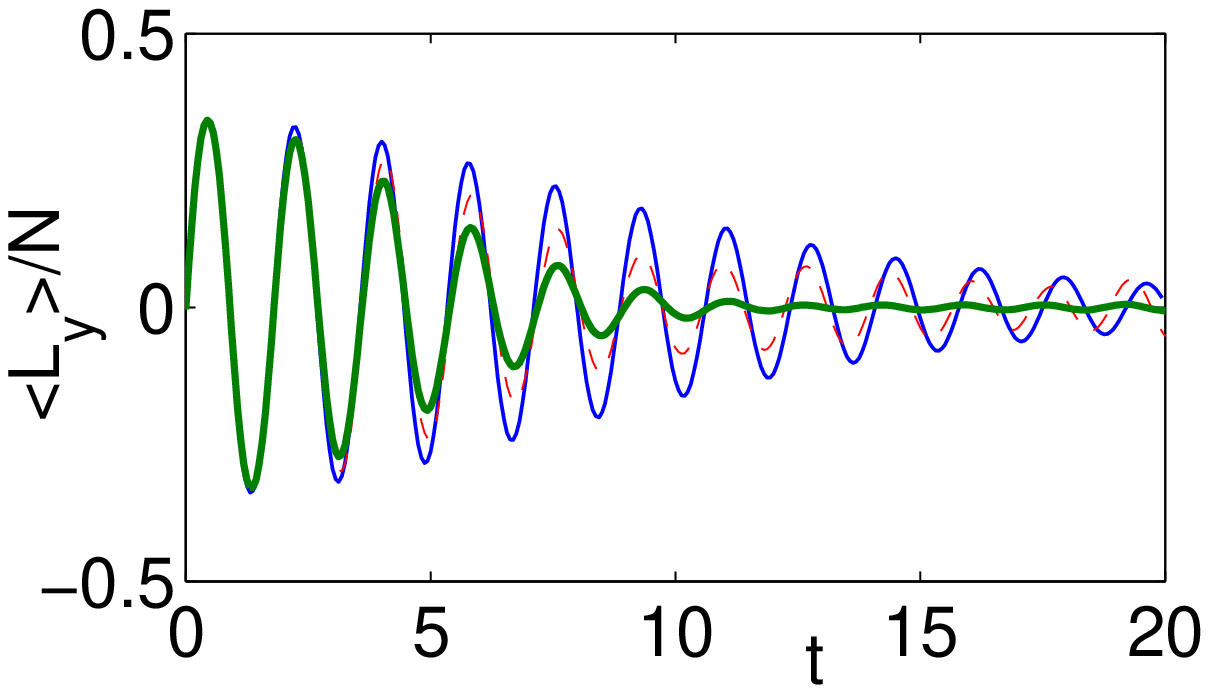}
\caption{\label{fig-qens2}
(Color online)
Dephasing of an initially coherent state. 
Upper panel: Illustration of the dynamics of the $SU(2)$-Husimi function. 
Lower panel: Dynamics of the expectation value $\langle \hat L_y \rangle/N$ 
calculated exactly (solid blue line) and quasi-classically from
an $SU(2)$-Husimi distributed phase space ensemble (dashed red line)
and a Glauber-Husimi distributed ensemble (thick green line).}
\end{figure}

The Glauber coherent states (\ref{eqn-glauberstate}) can be decomposed 
as
\be
  \ket{\alpha_1,\alpha_2} = e^{-\frac{\alpha^2}{2}} \sum_{N=0}^\infty 
     \frac{\alpha^N e^{- i \chi N}}{\sqrt{N!}} \ket{p,q}_N \, .
\ee
into $SU(2)$ coherent states with different particle numbers $N$,
where the respective parameters are related by
\be
  \alpha_1 = \alpha e^{i \chi} \sqrt{p}    \quad \mbox{and} \quad
  \alpha_2 = \alpha e^{i \chi} \sqrt{1-p} \, e^{-iq}.
\ee
Thus, $\alpha$ is the amplitude and $\chi$ is the global phase
of the Glauber coherent state. Using this decomposition one can 
easily relate the Husimi functions based on $SU(2)$ coherent states 
and Glauber states. Consider an arbitrary many-particle quantum 
state $\hat \rho$ with $N$ particles, which is represented by the $SU(2)$ 
Husimi function $Q_N(p,q)$. The Husimi function in terms of Glauber 
states is then given by
\be
   Q_{\rm Glauber}(\alpha_1,\alpha_2) 
   &=& \bra{\alpha_1,\alpha_2} \hat \rho \ket{\alpha_1,\alpha_2} \nn \\
   &=& \frac{e^{-\alpha^2} \alpha^{2N}}{N!} Q_N(p,q)
  \label{eqn-GN-Husimi}
\ee 
The physical implications of this mapping are clear: First of all, the 
Glauber-Husimi representation introduces artificial amplitude noise, since 
$|\alpha_1|^2 + |\alpha_2|^2 = \alpha^2$ is not fixed but Poisson-distributed.
The dynamical effects of this will be discussed below. 
The classical phase space is then given by 
$\mathbb{C}^2$ and the distribution functions thus depend on four variables.
In comparison, the approach based on the $SU(2)$ coherent states has some 
conceptual advantages as it directly embodies the conservation of the 
particle number $N$ and eliminates the global phase $\chi$. 
While the particle number is an operator in the $U(1)$-symmetry breaking approach,
it is a parameter in the $SU(2)$ mapping. Thus, the discussion of
the macroscopic limit is much simplified, as well as the estimation of the error, 
which vanishes as $\mathcal{O}(1/N)$.
Another approach discussing the convergence to the mean-field approximation
by an expansion in terms of the inverse particle number $1/N$ has been used 
in \cite{Cast97,Cast98}. 

The effects of the artificial amplitude fluctuations can be quite significant.
Let us illustrate them with a simple example shown in Fig.~\ref{fig-qens2}.
We consider the dynamics of an $SU(2)$ coherent state located at 
$(p,q) = (0.7,0)$ at $t=0$ for $\Delta=1$, $U=0.1$ and $N=50$ particles.
This state then rotates around the elliptic fixed point of the 
classical dynamics at $(p,q) = (0.5,0)$ and so the expectation 
values of the angular momentum operators (\ref{eqn-angular-op}) 
oscillate. These oscillations, however, are subject to dephasing 
and damping because of the quantum fluctuations of the initial state.
Fig.~\ref{fig-qens2} shows the dynamics of $s_y = \langle \hat L_y \rangle/N$ 
calculated exactly (solid blue line) and quasi-classically by the 
propagation of classical phase space ensembles. The ensembles were 
distributed according to the $SU(2)$ Husimi function (dashed red line) 
and the Glauber-Husimi function (\ref{eqn-GN-Husimi}) (thick green line).
While the damping is well reproduced by the $SU(2)$ ensemble, the artificial 
amplitude fluctuations of the Glauber-Husimi ensemble cause significant 
deviations. This is due to the fact that the oscillation frequency around 
the elliptic fixed point strongly depends on the effective nonlinearity 
$U\alpha$ and thus trajectories with different normalization $\alpha$
dephase rapidly.

Moreover, already the description of a BEC in terms of $U(1)$-symmetry 
breaking states reveals some ambiguities. For instance, consider an $SU(2)$ 
coherent state, representing a pure BEC. 
In the framework of the number-conserving phase space description,
the $P$-representation of this state is simply a delta function. 
On the contrary, the $P$-function in terms of Glauber coherent states is 
strongly singular, including higher-order derivatives of the delta-function. 
A pure BEC with a fixed number of particles is thus the most classical state 
within the $SU(2)$ phase space description whereas it is highly non-classical
in terms of the common Glauber phase space picture. In the first case an 
ensemble representation yields no dephasing at all while such a mapping 
is simply impossible in the latter case.

An extended and much more detailed overview over many different approaches 
to describe weakly-interacting condensates at finite, as well as at zero 
temperature is given in \cite{Prou08}, including symmetry-breaking methods,
as well as number conserving approaches. In this terminology, the method 
presented here can be seen as a combination of a stochastic phase space 
description with the number conserving approach.

\section{Conclusion and outlook}

In the present paper we have discussed the number-conserving phase
space description of small Bose-Hubbard systems and their generalization 
to an open quantum system. Apart from their theoretic value these systems 
have attracted considerable experimental interest within the last years,
especially the two-mode case and its open counterpart \cite{Albi05,Gati06}.

We have demonstrated the advantages of the (anti)-normal ordered quasi phase space
densities based on the $SU(M)$ coherent states instead of the common Glauber states
for the analysis and discussion of quantum states and especially of their dynamics. These 
representations allow a straightforward comparison of the many-body quantum system 
and its macroscopic counterpart given by the celebrated Gross-Pitaevskii equation.
For instance the quantum eigenstates localize on the classical phase space trajectories.
Moreover, we have presented an intuitive way to go beyond the usual mean-field limit:
Considering the macroscopic limit $N \rightarrow \infty$ with $g = UN$ without assuming a 
maximally localized state yields a classical Liouvillian flow for the time evolution
of the quasi-probability density or, equivalently, Monte Carlo ensembles of phase space 
trajectories. In contrast to the usual dynamics described by the GPE, this approach enables 
us to take into account arbitrary initial states and to estimate quantities depending 
on higher moments of the quantum state. Obviously this also holds for larger lattices,
for which a simulation of the full many-body dynamics is hard. 

As an example for the extended scope of application, we consider a BEC approaching a 
classically unstable fixed point. The condensate fraction rapidly decreases so that the 
dynamics cannot be described by a single GPE-trajectory any longer. This failure has
been denoted as the breakdown of the mean-field approximation in the literature 
\cite{Vard01b,Angl01}. However, this breakdown is resolved using the Liouville dynamics,
which provides a quite reasonable approximation for higher moments.
Thus, in contrast to the common mean-field approach, it can also be used to describe
and analyse chaotic systems. To underline this point, we study a driven two-mode system,
as well as the dynamics of the three-mode case, which features an interesting transition from
a regular to a mixed chaotic system, depending on the interaction strength.
A comparison to exact results shows an excellent agreement with the Liouville dynamics, 
while the common mean-field dynamics strongly diverge from the exact results and fail to 
describe the effects of the dynamical instability.
Moreover, the approximation of higher moments via the Liouville dynamics allows to investigate
many interesting quantities, like the single particle density matrix and the condensate fraction,
which are not accessible within the common approach. To underline this point, we study  
the depletion and the heating of the condensate.

In the last chapter, we give an overview over different methods, which can be used to
derive mean-field systems and their possible extensions and relate these to the approach 
presented here. Special attention is payed upon the comparison to methods based on 
$U(1)$-symmetry breaking Glauber states and the implications of the conservation of 
the total particle number. We show that even for simple examples the 
artificial noise in the particle number can result in larger deviations appearing on 
a much shorter time scale compared to the approach presented here.
Moreover, ambiguities in the description of macroscopic states and the set of problems 
considering the macroscopic limit with $\hat N$ being an operator can be tackled 
easily in the number conserving description where $N$ is simply a parameter. 
This completes the formal comparison of different possibilities of trial states in \cite{Fran00,Buon08}.
Note, however, that the benefits of a description based on generalized coherent states are not 
restricted to dynamical problems, see e.g. the semiclassical description of the ground 
state \cite{Buon05} and the excitation spectrum based on $SU(M)$ coherent states \cite{Fran00,Fran01}.

The introduction of generalized coherent states and the corresponding phase space 
distributions opens the door for the use of semiclassical methods for the analysis
of the dynamics of the Bose-Hubbard model. The Liouville approach is of course not 
capable to describe generic quantum effects such as tunneling in quantum phase space and 
(self-)interference. These features can be reconstructed using semiclassical 
coherent state propagators (see, e.g.~\cite{Ston00,Bara01} and references therein). 
In the first part of the present work \cite{07phase} we 
have introduced the phase space description and calculated the equations of motion 
for an arbitrary number of modes. The generalization of the ensemble method
to this case is straightforward and allows the approximate calculation of 
many-particle quantities such as the condensate fraction or higher moments
with small numerical efforts. Concrete applications to larger systems, as well as an 
extended study of the three-mode system will be subject of a future paper.

Moreover, the analysis of the mean-field dynamics has only recently been proven to 
be extremely useful for the understanding of the role of noise in two-mode Bose-Hubbard systems  
which face many theoretically and experimentally interesting features, like e.g. a 
stochastic resonance effect \cite{08mfdecay,08stores}. However, very little is 
known about the behaviour of larger dissipative systems, where the presented methods
could provide an illustrative and easily implementable tool.  

\begin{acknowledgments}
Support from the Studienstiftung des deutschen Volkes and the 
German Research Foundation (DFG) through the Graduiertenkolleg 792
and the research fellowship program (grant number WI 3415/1)
is gratefully acknowledged.
\end{acknowledgments}


\end{document}